\documentclass[twoside,11pt]{article}

\usepackage{jmlr2earxiv}
\usepackage{amsmath,amssymb,graphicx,algorithm,algorithmic,natbib,url}

\newcommand{\dunif}[0]{\mathrm{Uniform}}

\newcommand{\dnormal}[0]{\mathrm{Normal}}
\newcommand{\dt}[0]{\mathrm{t}}

\newcommand{\Exp}[0]{\textrm{Exponential}}

\newcommand{\ESS}[0]{\mathrm{ESS}}

\newcommand{\N}{\mathcal{N}}
\newcommand{\NEW}{\textrm{\tiny new}}
\newcommand{\OLD}{\textrm{\tiny old}}

\newcommand{\E}{\mathbb{E}}
\newcommand{\V}{\mathbb{V}}
\newcommand{\Eq}{\mathbb{E}_q}
\newcommand{\cL}{\mathcal{L}}
\newcommand{\cB}{\mathcal{B}}
\newcommand{\cC}{\mathcal{C}}

\newcommand{\LF}{\mathrm{Leapfrog}}

\ShortHeadings{The No-U-Turn Sampler}{Hoffman and Gelman} \firstpageno{1}

\begin{document}

\title{The No-U-Turn Sampler: Adaptively Setting Path Lengths
  in Hamiltonian Monte Carlo}

\author{\name Matthew D. Hoffman \email mdhoffma@cs.princeton.edu \\
       \addr Department of Statistics\\
       Columbia University\\
       New York, NY 10027, USA
       \AND
       \name Andrew Gelman \email gelman@stat.columbia.edu \\
       \addr Departments of Statistics and Political Science\\
       Columbia University\\
       New York, NY 10027, USA}

\maketitle

\begin{abstract}
Hamiltonian Monte Carlo (HMC) is a Markov chain Monte Carlo (MCMC)
algorithm that avoids the random walk behavior and sensitivity to
correlated parameters that plague many MCMC methods by taking a series
of steps informed by first-order gradient information. These features
allow it to converge to high-dimensional target distributions much
more quickly than simpler methods such as random walk Metropolis or
Gibbs sampling. However, HMC's performance is highly sensitive to two
user-specified parameters: a step size $\epsilon$ and a desired number
of steps $L$. In particular, if $L$ is too small then the algorithm
exhibits undesirable random walk behavior, while if $L$ is too large
the algorithm wastes computation.
We introduce the No-U-Turn Sampler (NUTS), an extension to HMC that
eliminates the need to set a number of steps $L$.  NUTS uses a
recursive algorithm to build a set of likely candidate points that
spans a wide swath of the target distribution, stopping automatically
when it starts to double back and retrace its steps. Empirically, NUTS
perform at least as efficiently as and sometimes more efficiently than
a well tuned standard HMC method, without requiring user intervention
or costly tuning runs. We also derive a method for adapting the step
size parameter $\epsilon$ on the fly based on primal-dual averaging.
NUTS can thus be used with no hand-tuning at all. NUTS is also
suitable for applications such as BUGS-style automatic inference
engines that require efficient ``turnkey'' sampling algorithms.
\end{abstract}


\section{Introduction}
Hierarchical Bayesian models are a mainstay of the machine learning
and statistics communities. Exact posterior inference in such models
is rarely tractable, however, and so researchers and practitioners
must usually resort to approximate statistical inference methods.
Deterministic approximate inference algorithms (for example, those
reviewed by \citet{Wainwright:2008}) can be efficient, but introduce
bias and can be difficult to apply to some models. Rather than
computing a deterministic approximation to a target posterior (or
other) distribution, Markov chain Monte Carlo (MCMC) methods offer
schemes for drawing a series of correlated samples that will converge
in distribution to the target distribution \citep{Neal:1993}. MCMC
methods are sometimes less efficient than their deterministic
counterparts, but are more generally applicable and are asymptotically
unbiased.

Not all MCMC algorithms are created equal. For complicated models with
many parameters, simple methods such as random-walk Metropolis
\citep{Metropolis:1953} and Gibbs sampling \citep{Geman:1984} may
require an unacceptably long time to converge to the target
distribution. This is in large part due to the tendency of these
methods to explore parameter space via inefficient random walks
\citep{Neal:1993}. When model parameters are continuous rather than
discrete, Hamiltonian Monte Carlo (HMC), also known as hybrid Monte
Carlo, is able to suppress such random walk behavior by means of a
clever auxiliary variable scheme that transforms the problem of
sampling from a target distribution into the problem of simulating
Hamiltonian dynamics \citep{Neal:2011}. The cost of HMC per
independent sample from a target distribution of dimension $D$ is
roughly $O(D^{5/4})$, which stands in sharp contrast with the $O(D^2)$
cost of random-walk Metropolis \citep{Creutz:1988}. 

HMC's increased efficiency comes at a price. First, HMC requires the
gradient of the log-posterior. Computing the gradient for a complex
model is at best tedious and at worst impossible, but this requirement
can be made less onerous by using automatic differentiation
\citep{Griewank:2008}.  Second, HMC requires that the user specify at
least two parameters: a step size $\epsilon$ and a number of steps $L$
for which to run a simulated Hamiltonian system. A poor choice of
either of these parameters will result in a dramatic drop in HMC's
efficiency. Methods from the adaptive MCMC literature (see
\citet{Andrieu:2008} for a review) can be used to tune $\epsilon$ on
the fly, but setting $L$ typically requires one or more costly tuning
runs, as well as the expertise to interpret the results of those
tuning runs. This hurdle limits the more widespread use of HMC, and
makes it challenging to incorporate HMC into a general-purpose
inference engine such as BUGS \citep{Gilks:1992}, JAGS
(http://mcmc-jags.sourceforge.net), Infer.NET \citep{infer.net},
HBC \citep{Daume:2007}, or PyMC \citep{Patil:2010}.

The main contribution of this paper is the No-U-Turn Sampler (NUTS),
an MCMC algorithm that closely resembles HMC, but eliminates the need
to choose the problematic number-of-steps parameter $L$. We also
provide a new dual averaging \citep{Nesterov:2009} scheme for
automatically tuning the step size parameter $\epsilon$ in both HMC
and NUTS, making it possible to run NUTS with no hand-tuning at
all. We will show that the tuning-free version of NUTS samples as
efficiently as (and sometimes more efficiently than) HMC, even
ignoring the cost of finding optimal tuning parameters for HMC. Thus,
NUTS brings the efficiency of HMC to users (and generic inference
systems) that are unable or disinclined to spend time tweaking an MCMC
algorithm.

\section{Hamiltonian Monte Carlo}
In Hamiltonian Monte Carlo (HMC) \citep{Neal:2011, Neal:1993,
  Duane:1987}, we introduce an auxiliary momentum variable $r_d$ for
each model variable $\theta_d$. In the usual implementation, these
momentum variables are drawn independently from the standard normal
distribution, yielding the (unnormalized) joint density
\begin{equation}
\textstyle
p(\theta, r) \propto \exp\{\cL(\theta) - \frac{1}{2}r \cdot r\},
\end{equation}
where $\cL$ is the logarithm of the joint density of the variables of
interest $\theta$ (up to a normalizing constant) and $x\cdot y$
denotes the inner product of the vectors $x$ and $y$.  We can
interpret this augmented model in physical terms as a fictitious
Hamiltonian system where $\theta$ denotes a particle's position in
$D$-dimensional space, $r_d$ denotes the momentum of that particle in
the $d$th dimension, $\cL$ is a position-dependent negative potential
energy function, $\frac{1}{2}r \cdot r$ is the kinetic energy of the particle,
and $\log p(\theta, r)$ is the negative energy of the particle. We can
simulate the evolution over time of the Hamiltonian dynamics of this
system via the ``leapfrog'' integrator, which proceeds according to
the updates
\begin{equation}
\label{eq:leapfrog}
r^{t + \epsilon/2} = r^t + (\epsilon/2) \nabla_\theta \cL(\theta^t);
\quad \theta^{t + \epsilon} = \theta^t + \epsilon r^{t+\epsilon/2};
\quad r^{t + \epsilon} = r^{t+\epsilon/2} + (\epsilon/2)\nabla_\theta
\cL(\theta^{t+\epsilon}),
\end{equation}
where $r^t$ and $\theta^t$ denote the values of the momentum and
position variables $r$ and $\theta$ at time $t$ and $\nabla_\theta$
denotes the gradient with respect to $\theta$. Since the update for
each coordinate depends only on the other coordinates, the leapfrog
updates are volume-preserving---that is, the volume of a region
remains unchanged after mapping each point in that region to a new
point via the leapfrog integrator.


\begin{algorithm}[tb]
   \caption{Hamiltonian Monte Carlo}
   \label{alg:HMC}
\begin{algorithmic}
  \STATE Given $\theta^0$, $\epsilon$, $L$, $\cL, M$:
  \FOR{$m=1$ to $M$}
  \STATE Sample $r^0\sim\N(0, I)$.
  \STATE Set $\theta^m\leftarrow \theta^{m-1}, \tilde\theta\leftarrow \theta^{m-1}, \tilde r\leftarrow r^0$.
  \FOR{$i=1$ to $L$}
  \STATE Set $\tilde\theta, \tilde r \leftarrow \LF(\tilde\theta, \tilde r, \epsilon)$.
  \ENDFOR
  \STATE With probability $\alpha = \min\left\{1, \frac{\exp\{\cL(\tilde\theta)-\frac{1}{2}\tilde r\cdot\tilde r\}}
         {\exp\{\cL(\theta^{m-1}) - \frac{1}{2} r^{0}\cdot r^{0}\}}\right\},$
  set $\theta^m\leftarrow\tilde \theta$, $r^m\leftarrow -\tilde r$.
  \ENDFOR
  \STATE
  \STATE {\bf function} $\LF(\theta, r, \epsilon)$
  \STATE Set $\tilde r \leftarrow r +
  (\epsilon/2)\nabla_\theta\cL(\theta)$.
  \STATE Set $\tilde \theta \leftarrow \theta +
  \epsilon \tilde r$.
  \STATE Set $\tilde r \leftarrow \tilde r +
  (\epsilon/2)\nabla_\theta\cL(\tilde\theta)$.
  \RETURN $\tilde \theta, \tilde r$.
\end{algorithmic}
\end{algorithm}


A standard procedure for drawing $M$ samples via Hamiltonian Monte
Carlo is described in Algorithm \ref{alg:HMC}. $I$ denotes the
identity matrix and $\N(\mu, \Sigma)$ denotes a multivariate normal
distribution with mean $\mu$ and covariance matrix $\Sigma$. For each
sample $m$, we first resample the momentum variables from a standard
multivariate normal, which can be inetpreted as a Gibbs sampling
update. We then apply $L$ leapfrog updates to the position and
momentum variables $\theta$ and $r$, generating a proposal
position-momentum pair $\tilde\theta, \tilde r$. We propose setting
$\theta^m=\tilde\theta$ and $r^m=-\tilde r$, and accept or reject this
proposal according to the Metropolis algorithm
\citep{Metropolis:1953}. This is a valid Metropolis proposal because
it is time-reversible and the leapfrog integrator is
volume-preserving; using an algorithm for simulating Hamiltonian
dynamics that did not preserve volume would seriously complicate the
computation of the Metropolis acceptance probability. The negation of
$\tilde r$ in the proposal is theoretically necessary to produce
time-reversibility, but can be omitted in practice if one is only
interested in sampling from $p(\theta)$. The algorithm's original
name, ``Hybrid Monte Carlo,'' refers to the hybrid approach of
alternating between updating $\theta$ and $r$ via Hamiltonian
simulation and updating $r$ via Gibbs sampling.

The term $\log \frac{p(\tilde\theta, \tilde r)}{p(\theta, r)}$, on
which the acceptance probability $\alpha$ depends, is the negative
change in energy of the simulated Hamiltonian system from time 0 to
time $\epsilon L$. If we could simulate the Hamiltonian dynamics
exactly, then $\alpha$ would always be 1, since energy is conserved in
Hamiltonian systems. The error introduced by using a discrete-time
simulation depends on the step size parameter
$\epsilon$---specifically, the change in energy $|\log
\frac{p(\tilde\theta, \tilde r)}{p(\theta, r)}|$ is proportional to
$\epsilon^2$ for large $L$, or $\epsilon^3$ if $L=1$
\citep{Leimkuhler:2004}. In theory the error can grow without bound as
a function of $L$, but in practice it typically does not when using
the leapfrog discretization. This allows us to run HMC with many
leapfrog steps, generating proposals for $\theta$ that have high
probability of acceptance even though they are distant from the
previous sample.

The performance of HMC depends strongly on choosing suitable values
for $\epsilon$ and $L$.
If $\epsilon$ is too large, then the simulation will be inaccurate and
yield low acceptance rates. If $\epsilon$ is too small, then
computation will be wasted taking many small steps. If $L$ is too
small, then successive samples will be close to one another, resulting
in undesirable random walk behavior and slow mixing. If $L$ is too
large, then HMC will generate trajectories that loop back and retrace
their steps. This is doubly wasteful, since work is being done to
bring the proposal $\tilde \theta$ {\it closer} to the initial
position $\theta^{m-1}$. Worse, if $L$ is chosen so that the
parameters jump from one side of the space to the other each
iteration, then the Markov chain may not even be ergodic
\citep{Neal:2011}. More realistically, an unfortunate choice of $L$
may result in a chain that is ergodic but slow to move between regions
of low and high density.

\section{Eliminating the Need to Hand-Tune HMC}
HMC is a powerful algorithm, but its usefulness is limited by the need
to tune the step size parameter $\epsilon$
and number of steps $L$. Tuning these parameters for any particular
problem requires some expertise, and usually one or more preliminary
runs. Selecting $L$ is particularly problematic; it is difficult to
find a simple metric for when a trajectory is too short, too long, or
``just right,'' and so practitioners commonly rely on heuristics based
on autocorrelation statistics from preliminary runs \citep{Neal:2011}.

Below, we present the No-U-Turn Sampler (NUTS), an extension of HMC
that eliminates the need to specify a fixed value of $L$. In section
\ref{sec:sa} we present schemes for setting $\epsilon$ based on the
dual averaging algorithm of \citet{Nesterov:2009}.

\subsection{No-U-Turn Hamiltonian Monte Carlo}
\label{sec:NUTS}
Our first goal is to devise an MCMC sampler that retains HMC's ability
to suppress random walk behavior without the need to set the number
$L$ of leapfrog steps that the algorithm takes to generate a proposal.
We need some criterion to tell us when we have simulated the dynamics
for ``long enough,'' i.e., when running the simulation for more steps
would no longer increase the distance between the proposal
$\tilde\theta$ and the initial value of $\theta$. We use a convenient
criterion based on the dot product between $\tilde r$ (the current
momentum) and $\tilde \theta- \theta$ (the vector from our initial
position to our current position), which is the derivative with
respect to time (in the Hamiltonian system) of half the squared
distance between the initial position $\theta$ and the current
position $\tilde \theta$:
\begin{equation}
\label{eq:timederiv}
\frac{d}{dt} \frac{(\tilde\theta - \theta)\cdot(\tilde\theta - \theta)}{2}
= (\tilde\theta - \theta) \cdot \frac{d}{dt} (\tilde\theta - \theta)
= (\tilde\theta - \theta) \cdot \tilde r.
\end{equation}
In other words, if we were to run the simulation for an infinitesimal
amount of additional time, then this quantity is proportional to the
progress we would make away from our starting point $\theta$.

This suggests an algorithm in which one runs leapfrog steps until the
quantity in equation \ref{eq:timederiv} becomes less than 0; such an
approach would simulate the system's dynamics until the proposal
location $\tilde\theta$ started to move back towards
$\theta$. Unfortunately this algorithm does not guarantee time
reversibility, and is therefore not guaranteed to converge to the
correct distribution. NUTS overcomes this issue by means of a
recursive algorithm reminiscent of the doubling procedure devised by
\citet{Neal:2003} for slice sampling.

\begin{figure}[t]
\begin{center}
  \centerline{\includegraphics[width=1\columnwidth]{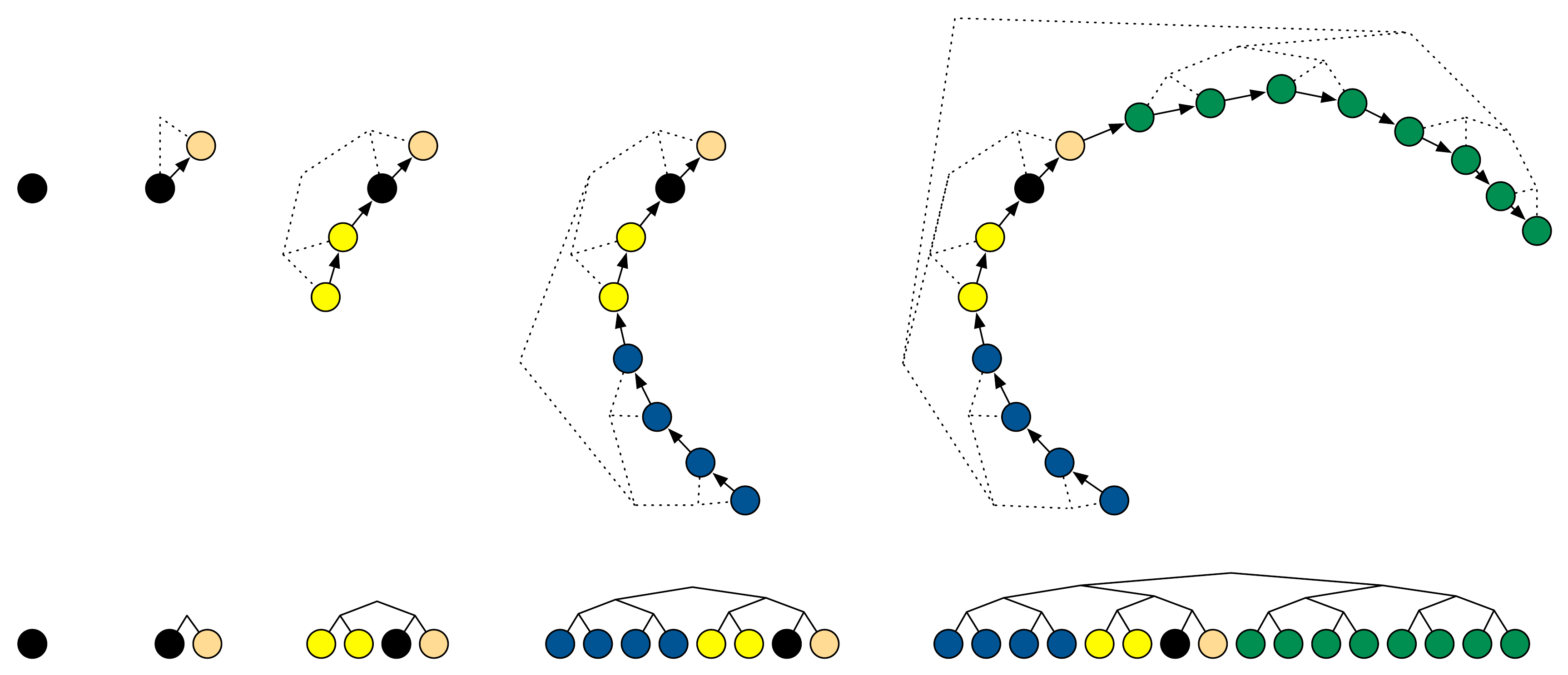}}
\end{center}
\vskip -0.3in
\caption{Example of building a binary tree via repeated doubling.
  Each doubling proceeds by choosing a direction (forwards or
  backwards in time) uniformly at random, then simulating Hamiltonian
  dynamics for $2^j$ leapfrog steps in that direction, where $j$ is
  the number of previous doublings (and the height of the binary
  tree). The figures at top show a trajectory in two dimensions (with
  corresponding binary tree in dashed lines) as it evolves over four
  doublings, and the figures below show the evolution of the binary
  tree. In this example, the directions chosen were forward (light
  orange node), backward (yellow nodes), backward (blue nodes), and
  forward (green nodes).}
\vskip -0.1in
\label{fig:doubling}
\end{figure}

\begin{figure}[t]
\begin{center}
  \vskip -0.3in
  \centerline{\includegraphics[width=1.125\columnwidth]{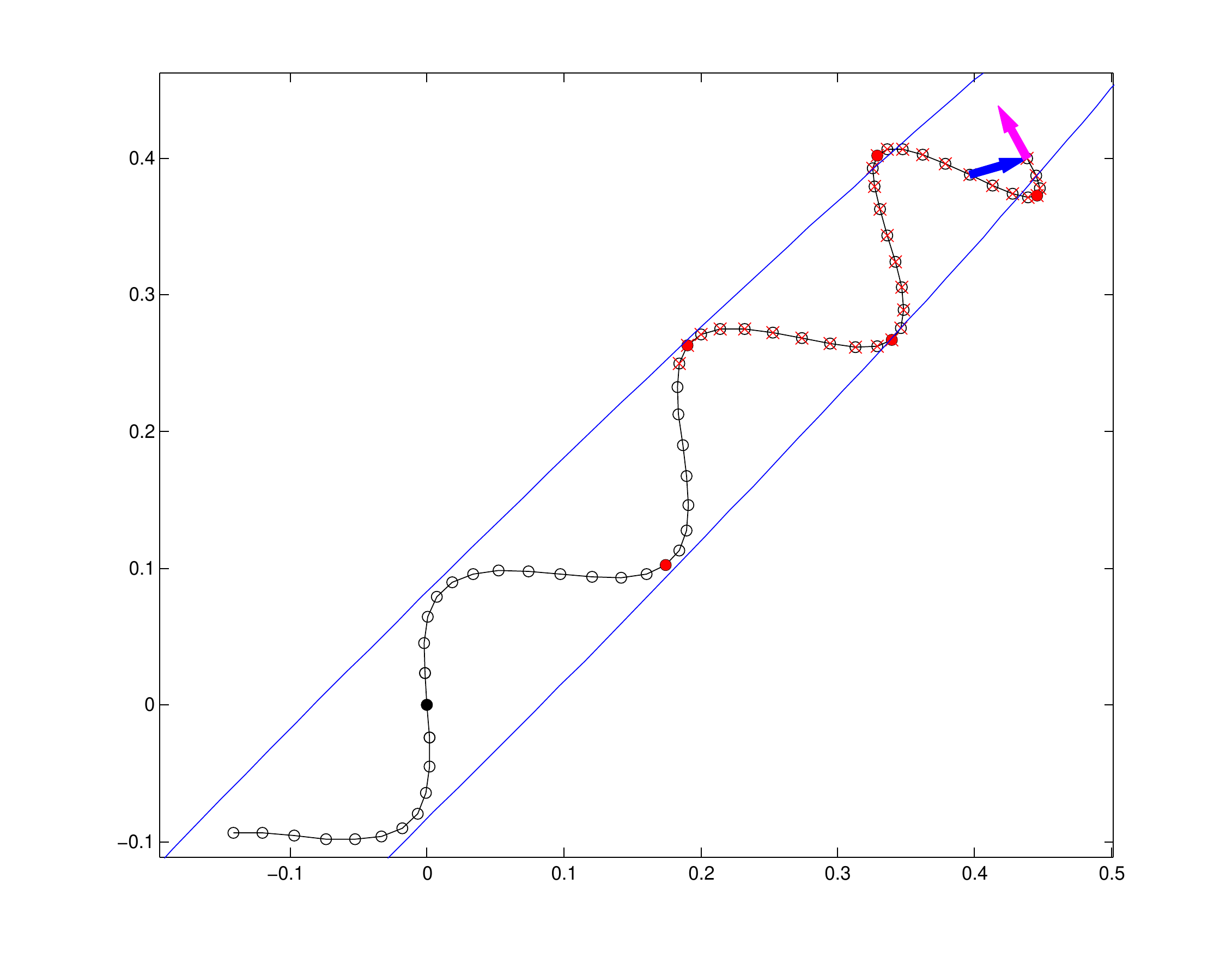}}
  \vskip -0.3in
\end{center}
\vskip -0.3in
\caption{Example of a trajectory generated during one iteration of
  NUTS. The blue ellipse is a contour of the target distribution, the
  black open circles are the positions $\theta$ traced out by the
  leapfrog integrator and associated with elements of the set of
  visited states $\cB$, the black solid circle is the starting
  position, the red solid circles are positions associated with states
  that must be excluded from the set $\cC$ of possible next samples
  because their joint probability is below the slice variable $u$, and
  the positions with a red ``x'' through them correspond to states
  that must be excluded from $\cC$ to satisfy detailed balance. The
  blue arrow is the vector from the positions associated with the
  leftmost to the rightmost leaf nodes in the rightmost height-3
  subtree, and the magenta arrow is the (normalized) momentum vector
  at the final state in the trajectory. The doubling process stops
  here, since the blue and magenta arrows make an angle of more than
  90 degrees. The crossed-out nodes with a red ``x'' are in the right
  half-tree, and must be ignored when choosing the next sample.}
\vskip -0.1in
\label{fig:stopping}
\end{figure}

NUTS begins by introducing a slice variable $u$ with conditional
distribution $p(u|\theta,
r)=\dunif(u;[0,\exp\{\cL(\theta)-\frac{1}{2}r\cdot r\}])$, which
renders the conditional distribution
$p(\theta,r|u)=\dunif(\theta,r;\{\theta',
r'|\exp\{\cL(\theta)-\frac{1}{2}r\cdot r\}\ge u\})$.
This slice sampling step is not strictly necessary, but it simplifies
both the derivation and the implementation of NUTS. In addition to
being more complicated, the analogous algorithm that eliminates the
slice variable seems empirically to be slightly less efficient than
the algorithm presented in this paper.

At a high level, after resampling $u|\theta, r$, NUTS uses the
leapfrog integrator to trace out a path forwards and backwards in
fictitious time, first running forwards or backwards 1 step, then
forwards or backwards 2 steps, then forwards or backwards 4 steps,
etc. This doubling process implicitly builds a balanced binary tree
whose leaf nodes correspond to position-momentum states, as
illustrated in Figure \ref{fig:doubling}. The doubling is halted when
the subtrajectory from the leftmost to the rightmost nodes of any
balanced subtree of the overall binary tree starts to double back on
itself (i.e., the fictional particle starts to make a ``U-turn'').
At this point NUTS stops the simulation and samples from among the set
of points computed during the simulation, taking care to preserve
detailed balance. Figure \ref{fig:stopping} illustrates an example of
a trajectory computed during an iteration of NUTS. 

Pseudocode implementing a efficient version of NUTS is provided in
Algorithm \ref{alg:NUTS}. A detailed derivation follows below, along
with a simplified version of the algorithm that motivates and builds
intuition about Algorithm \ref{alg:NUTS} (but uses much more memory
and makes smaller jumps).

\subsubsection{Derivation of simplified NUTS algorithm}
NUTS further augments the model
$p(\theta,r)\propto\exp\{\cL(\theta)-\frac{1}{2}r\cdot r\}$ with a
slice variable $u$ \citep{Neal:2003}.  The joint probability of
$\theta, r,$ and $u$ is
\begin{equation}
\label{eq:augmented}
\textstyle
p(\theta, r, u) \propto \mathbb{I}[u\in[0,
\exp\{\cL(\theta) - \frac{1}{2}r\cdot r
\}]],
\end{equation}
where $\mathbb{I}[\cdot]$ is 1 if the expression in brackets is true
and 0 if it is false. The (unnormalized) marginal probability of
$\theta$ and $r$ (integrating over $u$) is
\begin{equation}
\textstyle p(\theta, r) \propto \exp\{\cL(\theta) - \frac{1}{2}r\cdot r\},
\end{equation}
as in standard HMC. The conditional probabilities $p(u|\theta, r)$ and
$p(\theta, r|u)$ are each uniform, so long as the condition
$u\le\exp\{\cL(\theta)-\frac{1}{2}r\cdot r\}$ is satisfied.

We also add a finite set $\mathcal{C}$ of candidate position-momentum
states and another finite set $\mathcal{B}\supseteq\mathcal{C}$ to the
model. $\cB$ will be the set of all position-momentum states that the
leapfrog integrator traces out during a given NUTS iteration, and
$\cC$ will be the subset of those states to which we can transition
without violating detailed balance. $\cB$ will be built up by randomly
taking forward and backward leapfrog steps, and $\cC$ will selected
deterministically from $\cB$. The random procedure for building $\cB$
and $\cC$ given $\theta,$ $r,$ $u,$ and $\epsilon$ will define a
conditional distribution $p(\cB, \cC | \theta, r, u, \epsilon)$, upon
which we place the following conditions:
\begin{enumerate}
\renewcommand{\labelenumi}{C.\arabic{enumi}:}
\item All elements of $\cC$ must be chosen in a way that preserves
  volume. That is, any deterministic transformations of $\theta, r$
  used to add a state $\theta', r'$ to $\cC$ must 
have a Jacobian with
  unit determinant.
\item $p((\theta, r)\in \cC | \theta, r, u, \epsilon)=1$.
\item $p(u \le \exp\{\cL(\theta') - \frac{1}{2}r'\cdot r' \} | 
  (\theta', r') \in \mathcal{C}) = 1$.
\item If $(\theta, r)\in \mathcal{C}$ and $(\theta', r')\in
  \mathcal{C}$ then for any $\cB$, $p(\mathcal{B}, \mathcal{C} |
  \theta, r, u, \epsilon)= p(\mathcal{B}, \mathcal{C} | \theta', r',
  u, \epsilon)$.
\end{enumerate}
C.1 ensures that $p(\theta, r|(\theta, r)\in\cC)\propto p(\theta, r)$,
i.e. if we restrict our attention to the elements of $\cC$ then we can
treat the unnormalized probability density of a particular element of
$\cC$ as an unnormalized probability mass.  C.2 says that the current
state $\theta, r$ must be included in $\mathcal{C}$. C.3 requires that
any state in $\mathcal{C}$ be in the slice defined by $u$, i.e., that
any state $(\theta', r')\in\mathcal{C}$ must have equal (and positive)
conditional probability density $p(\theta', r'|u)$.  C.4 states that
$\mathcal{B}$ and $\mathcal{C}$ must have equal probability of being
selected regardless of the current state $\theta, r$ as long as
$(\theta, r)\in\mathcal{C}$ (which it must be by C.2).

Deferring for the moment the question of how to construct and sample
from a distribution $p(\cB,\cC|\theta, r, u, \epsilon)$ that satisfies
these conditions, we will now show that the the following procedure
leaves the joint distribution $p(\theta, r, u, \mathcal{B},
\mathcal{C}|\epsilon)$ invariant:
\begin{enumerate}
\item sample $r\sim\N(0, I)$,
\item sample $u\sim\dunif([0, \exp\{\cL(\theta^t) -
  \frac{1}{2}r\cdot r \}])$,
\item sample $\mathcal{B}, \mathcal{C}$ from their
  conditional distribution $p(\cB,\cC | \theta^t,
  r, u, \epsilon)$,
\item sample $\theta^{t+1}, r\sim T(\theta^t, r, \mathcal{C})$,
\end{enumerate}
where $T(\theta', r' | \theta, r, \mathcal{C})$ is a
transition kernel that leaves the uniform distribution over
$\mathcal{C}$ invariant, i.e., $T$ must satisfy
\begin{equation}
\frac{1}{|\mathcal{C}|}\sum_{(\theta, r)\in\mathcal{C}} T(\theta',
r'|\theta, r, \mathcal{C}) = \frac{\mathbb{I}[(\theta', r')\in\mathcal{C}]}
{|\mathcal{C}|}
\end{equation}
for any $\theta', r'$. The notation $\theta^{t+1}, r\sim T(\theta^t,
r, \mathcal{C})$ denotes that we are resampling $r$ in a way that
depends on its current value.

Steps 1, 2, and 3 resample $r$, $u$, $\mathcal{B}$, and $\mathcal{C}$
from their conditional joint distribution given $\theta^t$, and
therefore together constitute a valid Gibbs sampling update. Step 4 is
valid because the joint distribution of $\theta$ and $r$ given $u,
\mathcal{B}, \mathcal{C}$, and $\epsilon$ is uniform on the elements
of $\mathcal{C}$:
\begin{equation}
\begin{split}
\label{eq:pthetargivenuc}
p(\theta, r | u, \mathcal{B}, \mathcal{C}, \epsilon)
& \propto p(\cB, \cC | \theta, r, u, \epsilon) p(\theta, r | u) \\
& \propto p(\cB, \cC | \theta, r, u, \epsilon) 
\textstyle \mathbb{I}[u\le\exp\{\cL(\theta) - \frac{1}{2}r\cdot r\}] \\
& \propto \mathbb{I}[(\theta, r)\in\cC].
\end{split}
\end{equation}
Condition C.1 allows us to treat the unnormalized conditional density
$p(\theta, r | u)\propto\mathbb{I}[u\le\exp\{\cL(\theta) -
  \frac{1}{2}r\cdot r\}]$ as an unnormalized conditional probability
mass function. Conditions C.2 and C.4 ensure that $p(\cB, \cC|\theta,
r, u, \epsilon)\propto\mathbb{I}[(\theta, r)\in\cC]$ because by C.2
$(\theta, r)$ must be in $\cC$, and by C.4 for any $\cB, \cC$ pair
$p(\cB, \cC|\theta, r, u, \epsilon)$ is constant as a function of
$\theta$ and $r$ as long as $(\theta, r)\in\cC$. Condition C.3 ensures
that $(\theta, r)\in\cC \Rightarrow u\le\exp\{\cL(\theta) -
\frac{1}{2}r\cdot r\}$ (so the $p(\theta, r | u, \epsilon)$ term is
redundant).
Thus, equation \ref{eq:pthetargivenuc} implies that the joint
distribution of $\theta$ and $r$ given $u$ and $\mathcal{C}$ is
uniform on the elements of $\mathcal{C}$, and we are free to choose a
new $\theta^{t+1}, r^{t+1}$ from any transition kernel that leaves
this uniform distribution on $\mathcal{C}$ invariant.

We now turn our attention to the specific form for $p(\cB, \cC |
\theta, r, u, \epsilon)$ used by NUTS. Conceptually, the generative
process for building $\cB$ proceeds by repeatedly doubling the size of
a binary tree whose leaves correspond to position-momentum
states. These states will constitute the elements of $\cB$. The
initial tree has a single node corresponding to the initial
state. Doubling proceeds by choosing a random direction
$v_j\sim\dunif(\{-1,1\})$ and taking $2^j$ leapfrog steps of size
$v_j\epsilon$ (i.e., forwards in fictional time if $v_j=1$ and
backwards in fictional time if $v_j=-1$), where $j$ is the current
height of the tree. (The initial single-node tree is defined to have
height 0.) For example, if $v_j=1$, the left half of the new tree is
the old tree and the right half of the new tree is a balanced binary
tree of height $j$ whose leaf nodes correspond to the $2^j$
position-momentum states visited by the new leapfrog trajectory. This
doubling process is illustrated in Figure \ref{fig:doubling}. Given
the initial state $\theta, r$ and the step size $\epsilon$, there
are $2^j$ possible trees of height $j$ that can be built according to
this procedure, each of which is equally likely. Conversely, the
probability of reconstructing a particular tree of height $j$ starting
from any leaf node of that tree is $2^{-j}$ regardless of which leaf
node we start from.

We cannot keep expanding the tree forever, of course. We want to
continue expanding $\cB$ until one end of the trajectory we are
simulating makes a ``U-turn'' and begins to loop back towards another
position on the trajectory. At that point continuing the simulation is
likely to be wasteful, since the trajectory will retrace its steps and
visit locations in parameter space close to those we have already
visited. We also want to stop expanding $\cB$ if the error in the
simulation becomes extremely large, indicating that any states
discovered by continuing the simulation longer are likely to have
astronomically low probability. (This may happen if we use a step size
$\epsilon$ that is too large, or if the target distribution includes
hard constraints that make the log-density $\cL$ go to $-\infty$ in
some regions.)

The second rule is easy to formalize---we simply stop doubling if the
tree includes a leaf node whose state $\theta, r$ satisfies
\begin{equation}
\label{eq:stoperror}
\cL(\theta)-\frac{1}{2}r\cdot r - \log u < -\Delta_\mathrm{max}
\end{equation}
for some nonnegative $\Delta_\mathrm{max}$. We recommend setting
$\Delta_\mathrm{max}$ to a large value like 1000 so that it does not
interfere with the algorithm so long as the simulation is even
moderately accurate.

We must be careful when defining the first rule so that we can build a
sampler that neither violates detailed balance nor introduces
excessive computational overhead. To determine whether to stop
doubling the tree at height $j$, NUTS considers the $2^j-1$ balanced
binary subtrees of the height-$j$ tree that have height greater than
0. NUTS stops the doubling process when for one of these subtrees the
states $\theta^-, r^-$ and $\theta^+, r^+$ associated with the
leftmost and rightmost leaves of that subtree satisfies
\begin{equation}
\label{eq:stopangle}
(\theta^+-\theta^-)\cdot r^- < 0 \quad\mathrm{or}\quad
(\theta^+-\theta^-)\cdot r^+ < 0.
\end{equation}
That is, we stop if continuing the simulation an infinitesimal amount
either forward or backward in time would reduce the distance between
the position vectors $\theta^-$ and $\theta^+$. Evaluating the
condition in equation \ref{eq:stopangle} for each balanced subtree of
a tree of height $j$ requires $2^{j+1}-2$ inner products, which is
comparable to the number of inner products required by the $2^j-1$
leapfrog steps needed to compute the trajectory. Except for very
simple models with very little data, the cost of these inner products
is usually negligible compared to the cost of computing gradients.

This doubling process defines a distribution $p(\cB|\theta, r, u,
\epsilon)$. We now define a deterministic process for deciding which
elements of $\cB$ go in the candidate set $\cC$, taking care to
satisfy conditions C.1--C.4 on $p(\cB,\cC|\theta, r, u, \epsilon)$
laid out above. C.1 is automatically satisfied, since leapfrog steps
are volume preserving and any element of $\cC$ must be within some
number of leapfrog steps of every other element of $\cC$. C.2 is
satisfied as long as we include the initial state $\theta, r$ in
$\cC$, and C.3 is satisfied if we exclude any element $\theta',
r'$ of $\cB$ for which $\exp\{\cL(\theta')-\frac{1}{2}r'\cdot r'\} <
u$. To satisfy condition C.4, we must ensure that $p(\cB,\cC|\theta,
r, u, \epsilon)=p(\cB,\cC|\theta', r', u, \epsilon)$ for any
$(\theta', r')\in\cC$.  For any start state $(\theta', r')\in\cB$,
there is at most one series of directions $\{v_0,\ldots,v_j\}$ for
which the doubling process will reproduce $\cB$, so as long as we
choose $\cC$ deterministically given $\cB$ either $p(\cB,\cC|\theta',
r', u, \epsilon)=2^{-j}=p(\cB,\cC|\theta, r, u, \epsilon)$ or
$p(\cB,\cC|\theta', r', u, \epsilon)=0$. Thus, condition C.4 will be
satisfied as long as we exclude from $\cC$ any state $\theta', r'$
that could not have generated $\cB$. The only way such a state can
arise is if starting from $\theta', r'$ results in the stopping
conditions in equations \ref{eq:stoperror} or \ref{eq:stopangle} being
satisfied before the entire tree has been built, causing the doubling
process to stop too early. There are two cases to consider:
\begin{enumerate}
\item The doubling procedure was stopped because either equation
  \ref{eq:stoperror} or equation \ref{eq:stopangle} was satisfied by a
  state or subtree added during the final doubling iteration. In this
  case we must exclude from $\cC$ any element of $\cB$ that was added
  during this final doubling iteration, since starting the doubling
  process from one of these would lead to a stopping condition being
  satisfied before the full tree corresponding to $\cB$ has been
  built. 
\item The doubling procedure was stopped because equation
  \ref{eq:stopangle} was satisfied for the leftmost and rightmost
  leaves of the full tree corresponding to $\cB$. In this case no
  stopping condition was met by any state or subtree until $\cB$ had
  been completed, and condition C.4 is automatically satisfied.
\end{enumerate}

\begin{algorithm}[t]
   \caption{Naive No-U-Turn Sampler}
   \label{alg:naive-NUTS}
\begin{algorithmic}
  \small
  \STATE Given $\theta^0$, $\epsilon$, $\cL$, $M$:
  \FOR{$m=1$ to $M$}
  \STATE Resample $r^0\sim\N(0, I)$.
  \STATE Resample $u \sim \dunif([0, \exp\{\cL(\theta^{m-1} - \frac{1}{2}r^0\cdot r^0\}])$
  \STATE Initialize $\theta^-= \theta^{m-1}$, $\theta^+= \theta^{m-1}$,
  $r^-= r^0$, $r^+= r^0$, $j= 0$,
  $\mathcal{C}= \{(\theta^{m-1}, r^0)\}, s= 1$.
  \WHILE {$s=1$}
  \STATE Choose a direction $v_j \sim \dunif(\{-1, 1\})$.
  \IF {$v_j = -1$}
  \STATE $\theta^-, r^-, -, -, \mathcal{C}', s' \leftarrow 
  \mathrm{BuildTree}(\theta^-, r^-, u, v_j, j, \epsilon)$.
  \ELSE
  \STATE $-, -, \theta^+, r^+, \mathcal{C}', s' \leftarrow 
  \mathrm{BuildTree}(\theta^+, r^+, u, v_j, j, \epsilon)$.
  \ENDIF
  \IF {$s' = 1$}
  \STATE $\mathcal{C} \leftarrow \mathcal{C} \cup \mathcal{C}'$.
  \ENDIF
  \STATE $s\leftarrow s' \mathbb{I}[(\theta^+-\theta^-)\cdot r^- \ge 0]
  \mathbb{I}[(\theta^+-\theta^-)\cdot r^+ \ge 0]$.
  \STATE $j\leftarrow j+1$.
  \ENDWHILE
  \STATE Sample $\theta^{m}, r$ uniformly at random from
  $\mathcal{C}$.
  \ENDFOR
  \STATE
  \STATE {\bf function} $\mathrm{BuildTree}(\theta, r, u, v, j,
  \epsilon)$
  \IF {$j=0$}
  \STATE {\it Base case---take one leapfrog step in the direction $v$.}
  \STATE $\theta', r'\leftarrow\LF(\theta, r, v\epsilon)$.
  \STATE $\mathcal{C}' \leftarrow \left\{\begin{array}{ll}
  \{(\theta', r')\} & \mbox{if $u\le\exp\{\cL(\theta')-\frac{1}{2}r'\cdot r'\}$}
  \\ \emptyset & \mbox{else} \end{array} \right.$
  \STATE $s' \leftarrow \mathbb{I}[u < \exp\{\Delta_\mathrm{max} + \cL(\theta')-\frac{1}{2}r'\cdot r'\}]$.
  \STATE {\bf return} $\theta', r', \theta', r', \mathcal{C}', s'$.
  \ELSE
  \STATE {\it Recursion---build the left and right subtrees.}
  \STATE $\theta^-, r^-, \theta^+, r^+, \mathcal{C}', s' \leftarrow
  \mathrm{BuildTree}(\theta, r, u, v, j-1, \epsilon)$.
  \IF {$v = -1$}
  \STATE $\theta^-, r^-, -, -, \mathcal{C}'', s'' \leftarrow
  \mathrm{BuildTree}(\theta^-, r^-, u, v, j-1, \epsilon)$.
  \ELSE
  \STATE $-, -, \theta^+, r^+, \mathcal{C}'', s'' \leftarrow
  \mathrm{BuildTree}(\theta^+, r^+, u, v, j-1, \epsilon)$.
  \ENDIF
  \STATE $s' \leftarrow s's'' \mathbb{I}[(\theta^+-\theta^-)\cdot r^- \ge 0]
  \mathbb{I}[(\theta^+-\theta^-)\cdot r^+ \ge 0]$.
  \STATE $\cC' \leftarrow \cC' \cup \cC''$.
  \STATE {\bf return} $\theta^-, r^-, \theta^+, r^+, \mathcal{C}', s'$.
  \ENDIF
\end{algorithmic}
\end{algorithm}

Algorithm \ref{alg:naive-NUTS} shows how to construct $\cC$
incrementally while building $\cB$. After resampling the initial
momentum and slice variables, it uses a recursive procedure resembling
a depth-first search that eliminates the need to explicitly store the
tree used by the doubling procedure. The $\mathrm{BuildTree}()$ function
takes as input an initial position $\theta$ and momentum $r$, a slice
variable $u$, a direction $v\in\{-1,1\}$, a depth $j$, and a step size
$\epsilon$. It takes $2^j$ leapfrog steps of size $v\epsilon$
(i.e. forwards in time if $v=1$ and backwards in time if $v=-1$), and
returns
\begin{enumerate}
\item the backwardmost and forwardmost position-momentum states
  $\theta^-, r^-$ and $\theta^+, r^+$ among the $2^j$ new states
  visited;
\item a set $\cC'$ of position-momentum states containing each
  newly visited state $\theta', r'$ for which
  $\exp\{\cL(\theta')-\frac{1}{2}r'\cdot r'\}>u$; and
\item an indicator variable $s$; $s=0$ indicates that a stopping
  criterion was met by some state or subtree of the subtree
  corresponding to the $2^j$ new states visited by
  $\mathrm{BuildTree}()$.
\end{enumerate}
At the top level, NUTS repeatedly calls $\mathrm{BuildTree}()$ to
double the number of points that have been considered until either
$\mathrm{BuildTree}()$ returns $s=0$ (in which case doubling stops and
the new set $\cC'$ that was just returned must be ignored) or equation
\ref{eq:stopangle} is satisfied for the new backwardmost and
forwardmost position-momentum states $\theta^-, r^-$ and $\theta^+,
r^+$ yet considered (in which case doubling stops but we can use the
new set $\cC'$). Finally, we select the next position and momentum
$\theta^{m}, r$ uniformly at random from $\cC$, the union of
all of the valid sets $\cC'$ that have been returned, which clearly
leaves the uniform distribution over $\cC$ invariant.

To summarize, Algorithm \ref{alg:naive-NUTS} defines a transition
kernel that leaves $p(\theta, r, u, \cB, \cC|\epsilon)$ invariant, and
therefore leaves the target distribution
$p(\theta)\propto\exp\{\cL(\theta)\}$ invariant. It does so by
resampling the momentum and slice variables $r$ and $u$, simulating a
Hamiltonian trajectory forwards and backwards in time until that
trajectory either begins retracing its steps or encounters a state
with very low probability, carefully selecting a subset $\cC$ of the
states encountered on that trajectory that lie within the slice
defined by the slice variable $u$, and finally choosing the next
position and momentum variables $\theta^{m}$ and $r$ uniformly at
random from $\cC$. Figure \ref{fig:stopping} shows an example of a
trajectory generated by an iteration of NUTS where equation
\ref{eq:stopangle} is satisfied by the height-3 subtree at the end of
the trajectory. Below, we will introduce some improvements to
algorithm \ref{alg:naive-NUTS} that boost the algorithm's memory
efficiency and allow it to make larger jumps on average.

\subsubsection{Efficient NUTS}
Algorithm \ref{alg:naive-NUTS} requires $2^{j}-1$ evaluations of
$\cL(\theta)$ and its gradient (where $j$ is the number of times
$\mathrm{BuildTree}()$ is called), and $O(2^j)$ additional operations to
determine when to stop doubling. In practice, for all but the smallest
problems the cost of computing $\cL$ and its gradient still dominates
the overhead costs, so the computational cost of algorithm
\ref{alg:naive-NUTS} per leapfrog step is comparable to that of a
standard HMC algorithm. However, Algorithm \ref{alg:naive-NUTS} also
requires that we store $2^j$ position and momentum vectors, which may
require an unacceptably large amount of memory. Furthermore, there are
alternative transition kernels that satisfy detailed balance with
respect to the uniform distribution on $\cC$ that produce larger jumps
on average than simple uniform sampling. Finally, if a stopping
criterion is satisfied in the middle of the final doubling iteration
then there is no point in wasting computation to build up a set $\cC'$
that will never be used.

The third issue is easily addressed---if we break out of the recursion
as soon as we encounter a zero value for the stop indicator $s$ then
the correctness of the algorithm is unaffected and we save some
computation. We can address the second issue by using a more
sophisticated transition kernel to move from one state $(\theta,
r)\in\cC$ to another state $(\theta', r')\in\cC$ while leaving the
uniform distribution over $\cC$ invariant. This kernel admits a
memory-efficient implementation that only requires that we store
$O(j)$ position and momentum vectors, rather than $O(2^j)$.

Consider the transition kernel
\begin{equation}
T(w'|w,\cC) = \left\{ \begin{array}{ll} 
\frac{\mathbb{I}[w'\in\cC^\NEW]}{|\cC^\NEW|} & 
\mbox{if $|\cC^\NEW|>|\cC^\OLD|$}, \\
\frac{|\cC^\NEW|}{|\cC^\OLD|}\frac{\mathbb{I}[w'\in\cC^\NEW]}{|\cC^\NEW|} 
+ \left(1-\frac{|\cC^\NEW|}{|\cC^\OLD|}\right)\mathbb{I}[w'=w] & 
\mbox{if $|\cC^\NEW|\le|\cC^\OLD|$}
\end{array} \right. ,
\end{equation}
where $w$ and $w'$ are shorthands for position-momentum states
$(\theta, r)$, $\cC^\NEW$ and $\cC^\OLD$ are disjoint subsets of $\cC$
such that $\cC^\NEW\cup\cC^\OLD=\cC$, and $w\in\cC^\OLD$. In English,
$T$ proposes a move from $\cC^\OLD$ to a random state in $\cC^\NEW$
and accepts the move with probability
$\frac{|\cC^\NEW|}{|\cC^\OLD|}$. This is equivalent to a
Metropolis-Hastings kernel with proposal distribution $q(w',
{\cC^\OLD}', {\cC^\NEW}' |w, \cC^\OLD,
\cC^\NEW)\propto\mathbb{I}[w'\in\cC^\NEW]\mathbb{I}[{\cC^\OLD}'=\cC^\NEW]
\mathbb{I}[{\cC^\NEW}'=\cC^\OLD]$, and it is straightforward to show
that it satisfies detailed balance with respect to the uniform
distribution on $\cC$, i.e.
\begin{equation}
p(w|\cC)T(w'|w,\cC) = p(w'|\cC)T(w|w',\cC),
\end{equation}
and that $T$ therefore leaves the uniform distribution over $\cC$
invariant. If we let $\cC^\NEW$ be the (possibly empty) set of
elements added to $\cC$ during the final iteration of the doubling
(i.e. those returned by the final call to $\mathrm{BuildTree}()$ and
$\cC^\OLD$ be the older elements of $\cC$, then we can replace the
uniform sampling of $\cC$ at the end of Algorithm \ref{alg:naive-NUTS}
with a draw from $T(\theta^t, r^t, \cC)$ and leave the uniform
distribution on $\cC$ invariant. In fact, we can apply $T$ after {\it
  every} doubling, proposing a move to each new half-tree in
turn. Doing so leaves the uniform distribution on each partially built
$\cC$ invariant, and therefore does no harm to the invariance of the
uniform distribution on the fully built set $\cC$.  Repeatedly
applying $T$ in this way increases the probability that we will jump
to a state $\theta^{t+1}$ far from the initial state $\theta^t$;
considering the process in reverse, it is as though we first tried to
jump to the other side of $\cC$, then if that failed tried to make a
more modest jump, and so on. This transition kernel is thus akin to
delayed-rejection MCMC methods \citep{Tierney:1999}, but in this
setting we can avoid the usual costs associated with evaluating new
proposals.

The transition kernel above still requires that we be able to sample
uniformly from the set $\cC'$ returned by $\mathrm{BuildTree}()$,
which may contain as many as $2^{j-1}$ elements. In fact, we can
sample from $\cC'$ without maintaining the full set $\cC'$ in memory
by exploiting the binary tree structure in Figure
\ref{fig:doubling}. Consider a subtree of the tree explored in a call
to $\mathrm{BuildTree}()$, and let $\cC_\mathrm{subtree}$ denote the
set of its leaf states that are in $\cC'$: we can factorize the
probability that a state $(\theta, r)\in\cC_\mathrm{subtree}$ will
be chosen uniformly at random from $\cC'$ as
\begin{gather}
p(\theta, r|\cC') = \frac{1}{|\cC'|} =
\frac{|\cC_\mathrm{subtree}|}{|\cC'|}\frac{1}{|\cC_\mathrm{subtree}|}
\\ \nonumber
= p((\theta, r)\in\cC_\mathrm{subtree} | \cC)
p(\theta, r | 
(\theta, r)\in\cC_\mathrm{subtree}, \cC).
\end{gather}
That is, $p(\theta, r|\cC')$ is the product of the probability of
choosing some node from the subtree multiplied by the probability of
choosing $\theta, r$ uniformly at random from
$\cC_\mathrm{subtree}$. We use this observation to sample from $\cC'$
incrementally as we build up the tree. Each subtree above the bottom
layer is built of two smaller subtrees. For each of these smaller
subtrees, we sample a $\theta, r$ pair from $p(\theta, r |
(\theta, r)\in\cC_\mathrm{subtree})$ to represent that subtree. We
then choose between these two pairs, giving the pair representing each
subtree weight proportional to how many elements of $\cC'$ are in that
subtree. This continues until we have completed the subtree associated
with $\cC'$ and we have returned a sample $\theta'$ from $\cC'$ and an
integer weight $n'$ encoding the size of $\cC'$, which is all we need
to apply $T$. This procedure only requires that we store $O(j)$
position and momentum vectors in memory, rather than $O(2^j)$, and
requires that we generate $O(2^j)$ extra random numbers (a cost that
again is usually very small compared with the $2^{j}-1$ gradient
computations needed to run the leapfrog algorithm).

Algorithm \ref{alg:NUTS} implements all of the above improvements in
pseudocode. Matlab code implementing the algorithm is also available
at \url{http://www.cs.princeton.edu/~mdhoffma}, and a C++
implementation will also be available as part of the
soon-to-be-released Stan inference package.

\subsection{Adaptively Tuning $\epsilon$}
\label{sec:sa}
Having addressed the issue of how to choose the number of steps $L$,
we now turn our attention to the step size parameter $\epsilon$. To
set $\epsilon$ for both NUTS and HMC, we propose using stochastic
optimization with vanishing adaptation \citep{Andrieu:2008},
specifically an adaptation of the primal-dual algorithm of
\citet{Nesterov:2009}.

\begin{algorithm}[t!]
   \caption{Efficient No-U-Turn Sampler}
   \label{alg:NUTS}
\begin{algorithmic}
  \small
  \STATE Given $\theta^0$, $\epsilon$, $\cL$, $M$:
  \FOR{$m=1$ to $M$}
  \STATE Resample $r^0\sim\N(0, I)$.
  \STATE Resample $u \sim \dunif([0, \exp\{\cL(\theta^{m-1} - \frac{1}{2}r^0\cdot r^0\}])$
  \STATE Initialize $\theta^-=\theta^{m-1}$, 
  $\theta^+=\theta^{m-1}$, $r^-=r^0, r^+=r^0, j=0, 
  \theta^{m}=\theta^{m-1}, n=1, s=1$.
  \WHILE {$s=1$}
  \STATE Choose a direction $v_j \sim \dunif(\{-1, 1\})$.
  \IF {$v_j = -1$}
  \STATE $\theta^-, r^-, -, -, \theta', n', s' \leftarrow 
  \mathrm{BuildTree}(\theta^-, r^-, u, v_j, j, \epsilon)$.
  \ELSE
  \STATE $-, -, \theta^+, r^+, \theta', n', s' \leftarrow 
  \mathrm{BuildTree}(\theta^+, r^+, u, v_j, j, \epsilon)$.
  \ENDIF
  \IF {$s' = 1$}
  \STATE With probability $\min\{1, \frac{n'}{n}\}$, set
  $\theta^{m}\leftarrow \theta'$.
  \ENDIF
  \STATE $n\leftarrow n + n'$.
  \STATE $s\leftarrow s' \mathbb{I}[(\theta^+-\theta^-)\cdot r^- \ge 0]
  \mathbb{I}[(\theta^+-\theta^-)\cdot r^+ \ge 0]$.
  \STATE $j\leftarrow j+1$.
  \ENDWHILE
  \ENDFOR
  \STATE
  \STATE {\bf function} $\mathrm{BuildTree}(\theta, r, u, v, j,
  \epsilon)$
  \IF {$j=0$}
  \STATE {\it Base case---take one leapfrog step in the direction $v$.}
  \STATE $\theta', r'\leftarrow\LF(\theta, r, v\epsilon)$.
  \STATE $n' \leftarrow\mathbb{I}[u\le\exp\{\cL(\theta')-\frac{1}{2}r'\cdot r'\}]$.
  \STATE $s' \leftarrow \mathbb{I}[u < \exp\{\Delta_\mathrm{max} + \cL(\theta')-\frac{1}{2}r'\cdot r'\}]$.
  \STATE {\bf return} $\theta', r', \theta', r', \theta', n', s'$.
  \ELSE
  \STATE {\it Recursion---implicitly build the left and right subtrees.}
  \STATE $\theta^-, r^-, \theta^+, r^+, \theta', n', s' \leftarrow
  \mathrm{BuildTree}(\theta, r, u, v, j-1, \epsilon)$.
  \IF {$s' = 1$}
  \IF {$v = -1$}
  \STATE $\theta^-, r^-, -, -, \theta'', n'', s'' \leftarrow
  \mathrm{BuildTree}(\theta^-, r^-, u, v, j-1, \epsilon)$.
  \ELSE
  \STATE $-, -, \theta^+, r^+, \theta'', n'', s'' \leftarrow
  \mathrm{BuildTree}(\theta^+, r^+, u, v, j-1, \epsilon)$.
  \ENDIF
  \STATE With probability $\frac{n''}{n'+n''}$, 
  set $\theta'\leftarrow\theta''$.
  \STATE $s' \leftarrow s'' \mathbb{I}[(\theta^+-\theta^-)\cdot r^- \ge 0]
  \mathbb{I}[(\theta^+-\theta^-)\cdot r^+ \ge 0]$
  \STATE $n' \leftarrow n' + n''$
  \ENDIF
  \STATE {\bf return} $\theta^-, r^-, \theta^+, r^+, \theta', n', s'$.
  \ENDIF
\end{algorithmic}
\end{algorithm}

Perhaps the most commonly used vanishing adaptation algorithm in MCMC
is the stochastic approximation method of
\citet{Robbins:1951}. Suppose we have a statistic $H_t$ that describes
some aspect of the behavior of an MCMC algorithm at iteration $t\ge
1$, and define its expectation $h( x)$ as
\begin{equation}
h( x)\equiv\E_t[H_t| x]\equiv
\lim_{T\rightarrow\infty}\frac{1}{T} \sum_{t=1}^T \E[H_t| x],
\end{equation}
where $ x\in\mathbb{R}$ is a tunable parameter to the MCMC
algorithm. For example, if $\alpha_t$ is the Metropolis acceptance
probability for iteration $t$, we might define $H_t =
\delta-\alpha_t$, where $\delta$ is the desired average acceptance
probability.  If $h$ is a nondecreasing function of $ x$ and a few
other conditions such as boundedness of the iterates $x_t$ are met
(see \citet{Andrieu:2008} for details), the update
\begin{equation}
 x_{t+1}\leftarrow x_t - \eta_t H_t
\end{equation}
is guaranteed to cause $h( x_t)$ to converge to 0 as long as the
step size schedule defined by $\eta_t$ satisfies the conditions
\begin{equation}
\label{eq:sasteps}
\sum_t\eta_t=\infty;\quad \sum_t\eta_t^2<\infty.
\end{equation}
These conditions are satisfied by schedules of the form $\eta_t\equiv
t^{-\kappa}$ for $\kappa\in(0.5, 1]$. As long as the per-iteration
  impact of the adaptation goes to 0 (as it will if $\eta_t\equiv
  t^{-\kappa}$ and $\kappa>0$) the asymptotic behavior of the sampler
  is unchanged. That said, in practice $ x$ often gets ``close
  enough'' to an optimal value well before the step size $\eta$ has
  gotten close enough to 0 to avoid disturbing the Markov chain's
  stationary distribution. A common practice is therefore to adapt any
  tunable MCMC parameters during the burn-in phase, and freeze the
  tunable parameters afterwards (e.g., \citep{Gelman:2004}).

\paragraph{Dual averaging:} 
The optimal values of the parameters to an MCMC algorithm during the
burn-in phase and the stationary phase are often quite different.
Ideally those parameters would therefore adapt quickly as we shift
from the sampler's initial, transient regime to its stationary
regime. However, the diminishing step sizes of Robbins-Monro give
disproportionate weight to the {\it early} iterations, which is the
opposite of what we want.

Similar issues motivate the dual averaging scheme of
\citet{Nesterov:2009}, an algorithm for nonsmooth and stochastic
convex optimization. Since solving an unconstrained convex
optimization problem is equivalent to finding a zero of a
nondecreasing function (i.e., the (sub)gradient of the cost function),
it is straightforward to adapt dual averaging to the problem of MCMC
adaptation by replacing stochastic gradients with the statistics
$H_t$. Again assuming that we want to find a setting of a parameter $
x\in\mathbb{R}$ such that $h( x)\equiv\E_t[H_t| x]=0$, we can apply
the updates
\begin{equation}
\label{eq:daupdates}
 x_{t+1}\leftarrow  \mu - 
\frac{\sqrt{t}}{\gamma}\frac{1}{t+t_0}\sum_{i=1}^t H_i;\quad
\bar x_{t+1}\leftarrow \eta_t x_{t+1} + (1-\eta_t)\bar x_t,
\end{equation}
where $\mu$ is a freely chosen point that the iterates $x_t$ are
shrunk towards, $\gamma>0$ is a free parameter that controls the
amount of shrinkage towards $ \mu$, $t_0\ge 0$ is a free parameter
that stabilizes the initial iterations of the algorithm, $\eta_t\equiv
t^{-\kappa}$ is a step size schedule obeying the conditions in
equation \ref{eq:sasteps}, and we define $\bar x_1 = x_1$. As in
Robbins-Monro, the per-iteration impact of these updates on $x$ goes
to 0 as $t$ goes to infinity. Specifically, for large $t$ we have
\begin{equation}
x_{t+1}-x_t = O(-H_t t^{-0.5}),
\end{equation}
which clearly goes to 0 as long as the statistic $H_t$ is bounded.
The sequence of averaged iterates $\bar x_t$ is guaranteed to converge
to a value such that $h(\bar x_t)$ converges to 0.

The update scheme in equation \ref{eq:daupdates} is slightly more
elaborate than the update scheme of \citet{Nesterov:2009}, which
implicitly has $t_0\equiv0$ and $\kappa\equiv 1$. Introducing these
parameters addresses issues that are more important in MCMC adaptation
than in more conventional stochastic convex optimization
settings. Setting $t_0>0$ improves the stability of the algorithm in
early iterations, which prevents us from wasting computation by trying
out extreme values.  This is particularly important for NUTS, and for
HMC when simulation lengths are specified in terms of the overall
simulation length $\epsilon L$ instead of a fixed number of steps
$L$. In both of these cases, lower values of $\epsilon$ result in more
work being done per sample, so we want to avoid casually trying out
extremely low values of $\epsilon$. Setting the parameter $\kappa<1$
allows us to give higher weight to more recent iterates and more
quickly forget the iterates produced during the early burn-in stages.
The benefits of introducing these parameters are less apparent in the
settings originally considered by Nesterov, where the cost of a
stochastic gradient computation is assumed to be constant and the
stochastic gradients are assumed to be drawn i.i.d. given the
parameter $x$.

Allowing $t_0>0$ and $\kappa\in(0.5,1]$ does not affect the asymptotic
  convergence of the dual averaging algorithm. For any $\kappa\in(0.5,
  1]$, $\bar x_t$ will eventually converge to the same value
    $\frac{1}{t}\sum_{i=1}^t x_t$.  We can rewrite the term
    $\frac{\sqrt{t}}{\gamma}\frac{1}{t+t_0}$ as
    $\frac{t\sqrt{t}}{\gamma (t+t_0)}\frac{1}{t}$;
    $\frac{t\sqrt{t}}{\gamma (t+t_0)}$ is still $O(\sqrt{t})$, which
    is the only feature needed to guarantee convergence.

We used the values $\gamma=0.05, t_0=10,$ and $\kappa=0.75$ for all
our experiments. We arrived at these values by trying a few settings
for each parameter by hand with NUTS and HMC (with simulation lengths
specified in terms of $\epsilon L$) on the stochastic volatility model
described below and choosing a value for each parameter that seemed to
produce reasonable behavior. Better results might be obtained with
further tweaking, but these default parameters seem to work
consistently well for both NUTS and HMC for all of the models that we
tested. It is entirely possible that these parameter settings may not
work as well for other sampling algorithms or for $H$ statistics other
than the ones described below.

\paragraph{Setting $\epsilon$ in HMC:}
In HMC we want to find a value for the step size $\epsilon$ that is
neither too small (which would waste computation by taking needlessly
tiny steps) nor too large (which would waste computation by causing
high rejection rates). A standard approach is to tune $\epsilon$ so
that HMC's average Metropolis acceptance probability is equal to some
value $\delta$. Indeed, it has been shown that (under fairly strong
assumptions) the optimal value of $\epsilon$ for a given simulation
length $\epsilon L$ is the one that produces an average Metropolis
acceptance probability of approximately 0.65 \citep{Beskos:2010,
  Neal:2011}. For HMC, we define a criterion
$h^{\mathrm{HMC}}(\epsilon)$ so that
\begin{equation}
H^{\mathrm{HMC}}_t \equiv \min\left\{1, \frac{p(\tilde \theta^t,
  \tilde r^t)}{p(\theta^{t-1}, r^{t,0})}\right\};
\quad h^{\mathrm{HMC}}(\epsilon) \equiv \E_t[H_t^{\mathrm{HMC}}|\epsilon], 
\end{equation}
where $\tilde \theta^t$ and $\tilde r^t$ are the proposed position and
momentum at the $t$th iteration of the Markov chain, $\theta^{t-1}$
and $r^{t,0}$ are the initial position and (resampled) momentum for
the $t$th iteration of the Markov chain, $H^{\mathrm{HMC}}_t$ is the
acceptance probability of this $t$th HMC proposal and
$h^{\mathrm{HMC}}$ is the expected average acceptance probability of
the chain in equilibrium for a fixed $\epsilon$. Assuming that
$h^{\mathrm{HMC}}$ is nonincreasing as a function of $\epsilon$, we
can apply the updates in equation \ref{eq:daupdates} with $H_t\equiv
\delta-H^{\mathrm{HMC}}_t$ and $x\equiv\log \epsilon$ to coerce
$h^{\mathrm{HMC}}=\delta$ for any $\delta\in(0,1)$.




\paragraph{Setting $\epsilon$ in NUTS:}
Since there is no single accept/reject step in NUTS we must define an
alternative statistic to Metropolis acceptance probability. For each
iteration we define the statistic $H^{\mathrm{NUTS}}_t$ and its
expectation when the chain has reached equilibrium as
\begin{equation}
H^{\mathrm{NUTS}}_t \equiv
\frac{1}{|\cB_t^\mathrm{final}|}
\sum_{\theta, r\in\cB_t^\mathrm{final}}
\min\left\{1,\frac{p(\theta,r)}{p(\theta^{t-1},r^{t,0})}\right\}
;\quad h^\mathrm{NUTS} \equiv \E_t[H_t^\mathrm{NUTS}],
\end{equation}
where $\cB_t^\mathrm{final}$ is the set of all states explored during
the final doubling of iteration $t$ of the Markov chain and
$\theta^{t-1}$ and $r^{t,0}$ are the initial position and (resampled)
momentum for the $t$th iteration of the Markov
chain. $H^\mathrm{NUTS}$ can be understood as the average acceptance
probability that HMC would give to the position-momentum states
explored during the final doubling iteration. As above, assuming that
$H^\mathrm{NUTS}$ is nonincreasing in $\epsilon$, we can apply the
updates in equation \ref{eq:daupdates} with $H_t\equiv
\delta-H^\mathrm{NUTS}$ and $x\equiv\log \epsilon$ to coerce
$h^\mathrm{NUTS}=\delta$ for any $\delta\in(0,1)$.

\begin{algorithm}[tb]
\caption{Heuristic for choosing an initial value of $\epsilon$}
\label{alg:initialepsilon}
\begin{algorithmic}
  \small
  \STATE {\bf function} $\mathrm{FindReasonableEpsilon}(\theta)$
  \STATE Initialize $\epsilon=1$, $r\sim\N(0,I)$.
  \STATE Set $\theta', r'\leftarrow \LF(\theta, r, \epsilon)$.
  \STATE $a\leftarrow 2\mathbb{I}\left[\frac{p(\theta', r')}{p(\theta, r)} > 0.5\right]-1.$
  \WHILE {$\left(\frac{p(\theta', r')}{p(\theta, r)}\right)^a > 2^{-a}$}
  \STATE $\epsilon\leftarrow 2^a \epsilon$.
  \STATE Set $\theta', r'\leftarrow \LF(\theta, r, \epsilon)$.
  \ENDWHILE
  \RETURN $\epsilon$.
\end{algorithmic}
\end{algorithm}

\begin{algorithm}[t!]
   \caption{Hamiltonian Monte Carlo with Dual Averaging}
   \label{alg:dacdhmc}
\begin{algorithmic}
  \small
  \STATE Given $\theta^0$, $\delta$, $\lambda$, $\cL, M, M^\mathrm{adapt}$:
  \STATE Set $\epsilon_0=\mathrm{FindReasonableEpsilon}(\theta), \mu=\log(10\epsilon_0), \bar\epsilon_0=1, \bar H_0=0,
  \gamma=0.05, t_0=10, \kappa=0.75.$
  \FOR{$m=1$ to $M$}
  \STATE Reample $r^0\sim\N(0, I)$.
  \STATE Set $\theta^m\leftarrow \theta^{m-1}, \tilde\theta\leftarrow \theta^{m-1}, \tilde r\leftarrow r^0,
  L_m=\max\{1,\mathrm{Round}(\lambda/\epsilon_{m-1})\}$.
  \FOR{$i=1$ to $L_m$}
  \STATE Set $\tilde\theta, \tilde r \leftarrow \LF(\tilde\theta, \tilde r, \epsilon_{m-1})$.
  \ENDFOR
  \STATE With probability $\alpha = \min\left\{1, \frac{\exp\{\cL(\tilde\theta)-\frac{1}{2}\tilde r\cdot\tilde r\}}
         {\exp\{\cL(\theta^{m-1}) - \frac{1}{2} r^{0}\cdot r^{0}\}}\right\},$
  set $\theta^m\leftarrow\tilde \theta, r^m\leftarrow -\tilde r$.
  \IF {$m \le M^\mathrm{adapt}$}
  \STATE Set $\bar H_m = \left(1-\frac{1}{m+t_0}\right)\bar H_{m-1} + \frac{1}{m+t_0}(\delta-\alpha)$.
  \STATE Set $\log\epsilon_{m} = \mu - \frac{\sqrt{m}}{\gamma}\bar H_m,
  \log\bar\epsilon_{m} = m^{-\kappa}\log\epsilon_{m} + (1-m^{-\kappa})\log\bar\epsilon_{m-1}.$
  \ELSE
  \STATE Set $\epsilon_{m} = \bar\epsilon_{M^\mathrm{adapt}}$.
  \ENDIF
  \ENDFOR
\end{algorithmic}
\end{algorithm}

\begin{algorithm}[]
   \caption{No-U-Turn Sampler with Dual Averaging}
   \label{alg:danuts}
\begin{algorithmic}
  \footnotesize
  \STATE Given $\theta^0$, $\delta$, $\cL, M, M^\mathrm{adapt}$:
  \STATE Set $\epsilon_0=\mathrm{FindReasonableEpsilon}(\theta), \mu=\log(10\epsilon_0), \bar\epsilon_0=1, \bar H_0=0,
  \gamma=0.05, t_0=10, \kappa=0.75.$
  \FOR{$m=1$ to $M$}
  \STATE Sample $r^0\sim\N(0, I)$.
  \STATE Resample $u \sim \dunif([0, \exp\{\cL(\theta^{m-1} - \frac{1}{2}r^{0}\cdot r^{0}\}])$
  \STATE Initialize $\theta^-=\theta^{m-1}$, 
  $\theta^+=\theta^{m-1}$, $r^-=r^{0}, r^+=r^{0}, j=0, 
  \theta^{m}=\theta^{m-1}, n=1, s=1$.
  \WHILE {$s=1$}
  \STATE Choose a direction $v_j \sim \dunif(\{-1, 1\})$.
  \IF {$v_j = -1$}
  \STATE $\theta^-, r^-, -, -, \theta', n', s', \alpha, n_\alpha \leftarrow 
  \mathrm{BuildTree}(\theta^-, r^-, u, v_j, j, \epsilon_{m-1} \theta^{m-1}, r^{0})$.
  \ELSE
  \STATE $-, -, \theta^+, r^+, \theta', n', s', \alpha, n_\alpha \leftarrow 
  \mathrm{BuildTree}(\theta^+, r^+, u, v_j, j, \epsilon_{m-1}, \theta^{m-1}, r^{0})$.
  \ENDIF
  \IF {$s' = 1$}
  \STATE With probability $\min\{1, \frac{n'}{n}\}$, set
  $\theta^{m}\leftarrow \theta'$.
  \ENDIF
  \STATE $n\leftarrow n + n'$.
  \STATE $s\leftarrow s' \mathbb{I}[(\theta^+-\theta^-)\cdot r^- \ge 0]
  \mathbb{I}[(\theta^+-\theta^-)\cdot r^+ \ge 0]$.
  \STATE $j\leftarrow j+1$.
  \ENDWHILE
  \IF {$m \le M^\mathrm{adapt}$}
  \STATE Set $\bar H_m = \left(1-\frac{1}{m+t_0}\right)\bar H_{m-1} + \frac{1}{m+t_0}(\delta-\frac{\alpha}{n_\alpha})$.
  \STATE Set $\log\epsilon_{m} = \mu - \frac{\sqrt{m}}{\gamma}\bar H_m,
  \log\bar\epsilon_{m} = m^{-\kappa}\log\epsilon_{m} + (1-m^{-\kappa})\log\bar\epsilon_{m-1}.$
  \ELSE
  \STATE Set $\epsilon_{m} = \bar\epsilon_{M^\mathrm{adapt}}$.
  \ENDIF
  \ENDFOR
  \STATE
  \STATE {\bf function} $\mathrm{BuildTree}(\theta, r, u, v, j,
  \epsilon, \theta^0, r^0)$
  \IF {$j=0$}
  \STATE {\it Base case---take one leapfrog step in the direction $v$.}
  \STATE $\theta', r'\leftarrow\LF(\theta, r, v\epsilon)$.
  \STATE $n' \leftarrow\mathbb{I}[u\le\exp\{\cL(\theta')-\frac{1}{2}r'\cdot r'\}]$.
  \STATE $s' \leftarrow \mathbb{I}[u < \exp\{\Delta_\mathrm{max} + \cL(\theta')-\frac{1}{2}r'\cdot r'\}]$.
  \STATE {\bf return} $\theta', r', \theta', r', \theta', n', s',
  \min\{1,\exp\{\cL(\theta')-\frac{1}{2}r'\cdot r' - \cL(\theta^0) + \frac{1}{2}r^0\cdot r^0\}\}, 1$.
  \ELSE
  \STATE {\it Recursion---implicitly build the left and right subtrees.}
  \STATE $\theta^-, r^-, \theta^+, r^+, \theta', n', s', \alpha', n_\alpha' \leftarrow
  \mathrm{BuildTree}(\theta, r, u, v, j-1, \epsilon, \theta^0, r^0)$.
  \IF {$s' = 1$}
  \IF {$v = -1$}
  \STATE $\theta^-, r^-, -, -, \theta'', n'', s'', \alpha'', n_\alpha'' \leftarrow
  \mathrm{BuildTree}(\theta^-, r^-, u, v, j-1, \epsilon, \theta^0, r^0)$.
  \ELSE
  \STATE $-, -, \theta^+, r^+, \theta'', n'', s'', \alpha'', n_\alpha'' \leftarrow
  \mathrm{BuildTree}(\theta^+, r^+, u, v, j-1, \epsilon, \theta^0, r^0)$.
  \ENDIF
  \STATE With probability $\frac{n''}{n'+n''}$, 
  set $\theta'\leftarrow\theta''$.
  \STATE Set $\alpha'\leftarrow\alpha' + \alpha''$, $n_\alpha'\leftarrow n_\alpha' + n_\alpha''$.
  \STATE $s' \leftarrow s'' \mathbb{I}[(\theta^+-\theta^-)\cdot r^- \ge 0]
  \mathbb{I}[(\theta^+-\theta^-)\cdot r^+ \ge 0]$
  \STATE $n' \leftarrow n' + n''$
  \ENDIF
  \STATE {\bf return} $\theta^-, r^-, \theta^+, r^+, \theta', n', s', \alpha', n_\alpha'$.
  \ENDIF
\end{algorithmic}
\end{algorithm}

\paragraph{Finding a good initial value of $\epsilon$:}
The dual averaging scheme outlined above should work for any initial
value $\epsilon_1$ and any setting of the shrinkage target
$\mu$. However, convergence will be faster if we start from a
reasonable setting of these parameters. We recommend choosing an
initial value $\epsilon_1$ according to the simple heuristic described
in Algorithm \ref{alg:initialepsilon}. In English, this heuristic
repeatedly doubles or halves the value of $\epsilon_1$ until the
acceptance probability of the Langevin proposal with step size
$\epsilon_1$ crosses 0.5. The resulting value of $\epsilon_1$ will
typically be small enough to produce reasonably accurate simulations
but large enough to avoid wasting large amounts of computation. We
recommend setting $\mu=\log(10\epsilon_1)$, since this gives the dual
averaging algorithm a preference for testing values of $\epsilon$ that
are larger than the initial value $\epsilon_1$. Large values of
$\epsilon$ cost less to evaluate than small values of $\epsilon$, and
so erring on the side of trying large values can save computation.


Algorithms \ref{alg:dacdhmc} and \ref{alg:danuts} show how to
implement HMC (with simulation length specified in terms of $\epsilon
L$ rather than $L$) and NUTS while incorporating the dual averaging
algorithm derived in this section, with the above initialization
scheme. Algorithm \ref{alg:dacdhmc} requires as input a target
simulation length $\lambda\approx\epsilon L$, a target mean acceptance
probability $\delta$, and a number of iterations $M^\mathrm{adapt}$
after which to stop the adaptation. Algorithm \ref{alg:danuts}
requires only a target mean acceptance probability $\delta$ and a
number of iterations $M^\mathrm{adapt}$. Matlab code implementing both
algorithms can be found at
\url{http://www.cs.princeton.edu/~mdhoffma}, and C++ implementations
will be available as part of the Stan inference package.

\section{Empirical Evaluation}
\label{sec:evaluation}
In this section we examine the effectiveness of the dual averaging
algorithm outlined in section \ref{sec:sa}, examine what values of the
target $\delta$ in the dual averaging algorithm yield efficient
samplers, and compare the efficiency of NUTS and HMC.

For each target distribution, we ran HMC (as implemented in algorithm
\ref{alg:dacdhmc}) and NUTS (as implemented in algorithm
\ref{alg:danuts}) with four target distributions for 2000 iterations,
allowing the step size $\epsilon$ to adapt via the dual averaging
updates described in section \ref{sec:sa} for the first 1000
iterations. In all experiments the dual averaging parameters were set
to $\gamma=0.05, t_0=10,$ and $\kappa=0.75$. We evaluated HMC with 10
logarithmically spaced target simulation lengths $\lambda$ per target
distribution. For each target distribution the largest value of
$\lambda$ that we tested was 40 times the smallest value of $\lambda$
that we tested, meaning that each successive $\lambda$ is
$40^{1/9}\approx 1.5$ times larger than the previous $\lambda$.  We
tried 15 evenly spaced values of the dual averaging target $\delta$
between 0.25 and 0.95 for NUTS and 8 evenly spaced values of the dual
averaging target $\delta$ between 0.25 and 0.95 for HMC. For each
sampler-simulation length-$\delta$-target distribution combination we
ran 10 iterations with different random seeds. In total, we ran 3,200
experiments with HMC and 600 experiments with NUTS.


We measure the efficiency of each algorithm in terms of effective
sample size (ESS) normalized by the number of gradient evaluations
used by each algorithm. The ESS of a set of $M$ correlated samples
$\theta^{1:M}$ with respect to some function $f(\theta)$ is the number
of independent draws from the target distribution $p(\theta)$ that
would give a Monte Carlo estimate of the mean under $p$ of $f(\theta)$
with the same level of precision as the estimate given by the mean of
$f$ for the correlated samples $\theta^{1:M}$. That is, the ESS of a
sample is a measure of how many independent samples a set of
correlated samples is worth for the purposes of estimating the mean of
some function; a more efficient sampler will give a larger ESS for
less computation. We use the number of gradient evaluations performed
by an algorithm as a proxy for the total amount of computation
performed; in all of the models and distributions we tested the
computational overhead of both HMC and NUTS is dominated by the cost
of computing gradients.  Details of the method we use to estimate ESS
are provided in appendix \ref{app:ess}. In each experiment, we
discarded the first 1000 samples as burn-in when estimating ESS.



ESS is inherently a univariate statistic, but all of the distributions
we test HMC and NUTS on are multivariate. Following
\citet{Girolami:2011} we compute ESS separately for each dimension and
report the minimum ESS across all dimensions, since we want our
samplers to effectively explore all dimensions of the target
distribution. For each dimension we compute ESS in terms of the
variance of the estimator of that dimension's mean and second central
moment (where the estimate of the mean used to compute the second
central moment is taken from a separate long run of 50,000 iterations
of NUTS with $\delta=0.5$), reporting whichever statistic has a lower
effective sample size. We include the second central moment as well as
the mean because for simulation lengths $\epsilon L$ that hit a
resonance of the target distribution HMC can produce samples that are
{\it anti-}correlated. These samples yield low-variance estimators of
parameter means, but very high-variance estimators of parameter
variances, so computing ESS only in terms of the mean of $\theta$ can
be misleading.

\subsection{Models and Datasets}
\label{sec:models}
To evaluate NUTS and HMC, we used the two algorithms to sample from
four target distributions, one of which was synthetic and the other
three of which are posterior distributions arising from real datasets.

\paragraph{250-dimensional multivariate normal (MVN):} 
In these experiments the target distribution was a zero-mean
250-dimensional multivariate normal with known precision matrix $A$,
i.e.,
\begin{equation}
\textstyle
p(\theta)\propto\exp\{-\frac{1}{2}\theta^T A \theta\}.
\end{equation}
The matrix $A$ was generated from a Wishart distribution with identity
scale matrix and 250 degrees of freedom. This yields a target
distribution with many strong correlations. The same matrix $A$ was
used in all experiments.

\paragraph{Bayesian logistic regression (LR):}
In these experiments the target distribution is the posterior of a
Bayesian logistic regression model fit to the German credit dataset
(available from the UCI repository \citep{Frank:2010}). The target
distribution is
\begin{equation}
\begin{split}
\textstyle
p(\alpha, \beta|x, y) & \propto p(y|x, \alpha, \beta)p(\alpha)p(\beta) \\
& \textstyle \propto \exp\{-\sum_i \log(1 + \exp\{-y_i (\alpha + x_i\cdot \beta\})
- \frac{1}{2\sigma^2}\alpha^2 - \frac{1}{2\sigma^2}\beta\cdot \beta\},
\end{split}
\end{equation}
where $x_i$ is a 24-dimensional vector of numerical predictors
associated with a customer $i$, $y_i$ is $-1$ if customer $i$ should
be denied credit and 1 if that customer should receive credit,
$\alpha$ is an intercept term, and $\beta$ is a vector of 24
regression coefficients. All predictors are normalized to have zero
mean and unit variance. $\alpha$ and each element of $\beta$ are given
weak zero-mean normal priors with variance $\sigma^2=100$. The dataset
contains predictor and response data for 1000 customers.

\label{sec:hconvergence}
\begin{figure}[t!]
\vskip-0.3in
\begin{center}
  \centerline{\includegraphics[width=1\columnwidth]{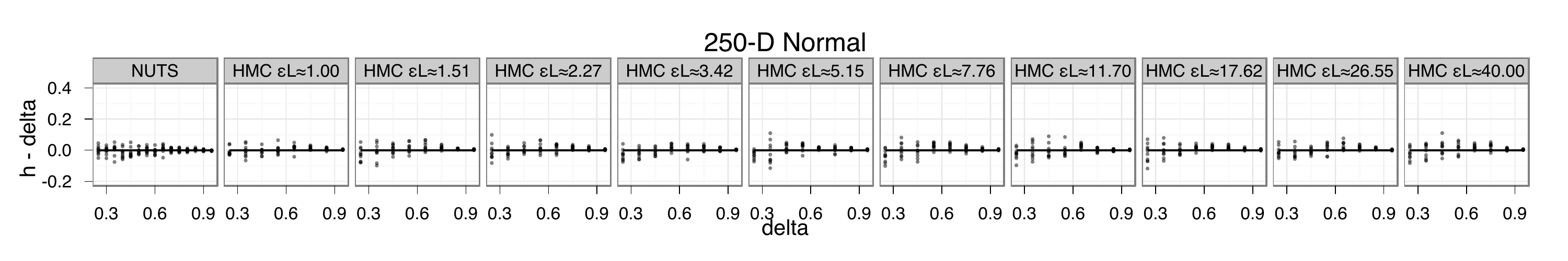}}
  \vskip-0.1in
  \centerline{\includegraphics[width=1\columnwidth]{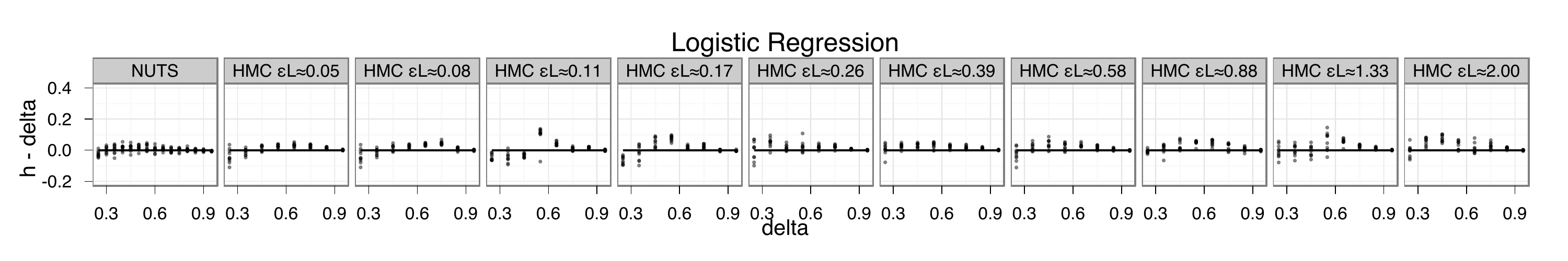}}
  \vskip-0.1in
  \centerline{\includegraphics[width=1\columnwidth]{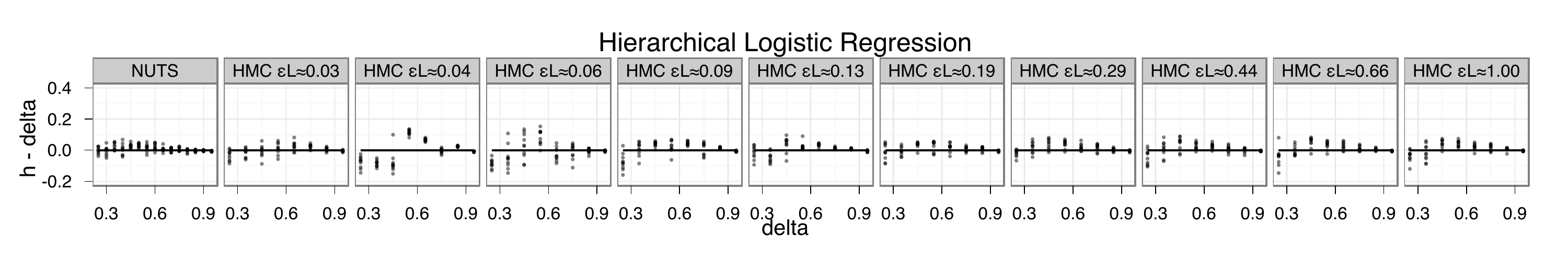}}
  \vskip-0.1in
  \centerline{\includegraphics[width=1\columnwidth]{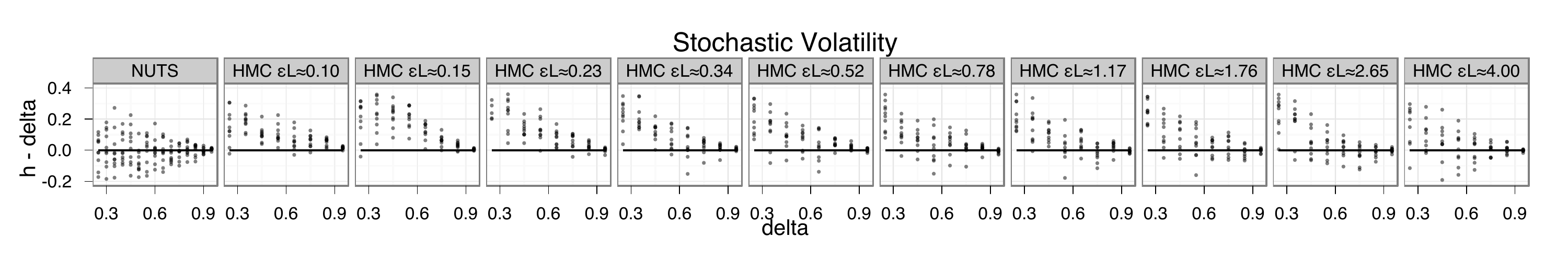}}
\end{center}
\vskip -0.3in
\caption{Discrepancies between the realized average acceptance
  probability statistic $h$ and its target $\delta$ for the
  multivariate normal, logistic regression, hierarchical logistic
  regression, and stochastic volatility models. Each point's distance
  from the x-axis shows how effectively the dual averaging algorithm
  tuned the step size $\epsilon$ for a single experiment. Leftmost
  plots show experiments run with NUTS, other plots show experiments
  run with HMC with a different setting of $\epsilon L$.}
\vskip -0.2in
\label{fig:hconvergence}
\end{figure}

\paragraph{Hierarchical Bayesian logistic regression (HLR):}
In these experiments the target distribution is again the posterior of
a Bayesian logistic regression model fit to the German credit dataset,
but this time the variance parameter in the prior on $\alpha$ and
$\beta$ is given an exponential prior and estimated as well. Also, we
expand the predictor vectors by including two-way interactions,
resulting in ${24 \choose 2} + 24 = 300$-dimensional vectors of
predictors $x$ and a 300-dimensional vector of coefficients
$\beta$. These elaborations on the model make for a more challenging
problem; the posterior is in higher dimensions, and the variance term
$\sigma^2$ interacts strongly with the remaining 301 variables.  The
target distribution for this problem is
\begin{equation}
\begin{split}
\textstyle
p(\alpha, \beta, \sigma^2|x, y) &\propto p(y|x, \alpha, \beta)p(\beta|\sigma^2)p(\alpha|\sigma^2)p(\sigma^2)
\\ &\textstyle\propto \exp\{-\sum_i \log(1 + \exp\{-y_i x_i\cdot \beta\})
- \frac{1}{2\sigma^2}\alpha^2 - \frac{1}{2\sigma^2}\beta\cdot \beta 
- \frac{N}{2}\log\sigma^2 - \lambda\sigma^2\},
\end{split}
\end{equation}
where $N=1000$ is the number of customers and $\lambda$ is the rate
parameter to the prior on $\sigma^2$.  We set $\lambda=0.01$, yielding
a weak exponential prior distribution on $\sigma^2$ whose mean and
standard deviation are 100.

\paragraph{Stochastic volatility (SV):}
In the final set of experiments the target distribution is the
posterior of a relatively simple stochastic volatility model fit to
3000 days of returns from the S\&P 500 index. The model assumes that
the observed values of the index are generated by the following
generative process:
\begin{gather}
\nonumber \tau\sim \Exp(100);\quad \nu\sim\Exp(100); \quad s_1\sim\Exp(100); \\
\textstyle \log s_{i>1}\sim \dnormal(\log s_{i-1}, \tau^{-1}); \quad
\frac{\log y_i - \log y_{i-1}}{s_i} \sim \dt_\nu,
\end{gather}
where $s_{i>1}$ refers to a scale parameter $s_i$ where $i>1$.  We
integrate out the precision parameter $\tau$ to speed mixing, leading
to the 3001-dimensional target distribution
\begin{multline}
\textstyle p(s, \nu|y)\propto e^{-0.01\nu}e^{-0.01s_1}
(\prod_{i=1}^{3000}\dt_\nu(s_i^{-1}(\log y_i - \log y_{i-1}))) \times \\
\textstyle 
(0.01 + 0.5 \sum_{i=2}^{3000} (\log s_i - \log s_{i-1})^2)^{-\frac{3001}{2}}.
\end{multline}

\subsection{Convergence of Dual Averaging}


\begin{figure}[t]
\vskip-0.3in
\begin{center}
  \centerline{\includegraphics[width=1.025\columnwidth]{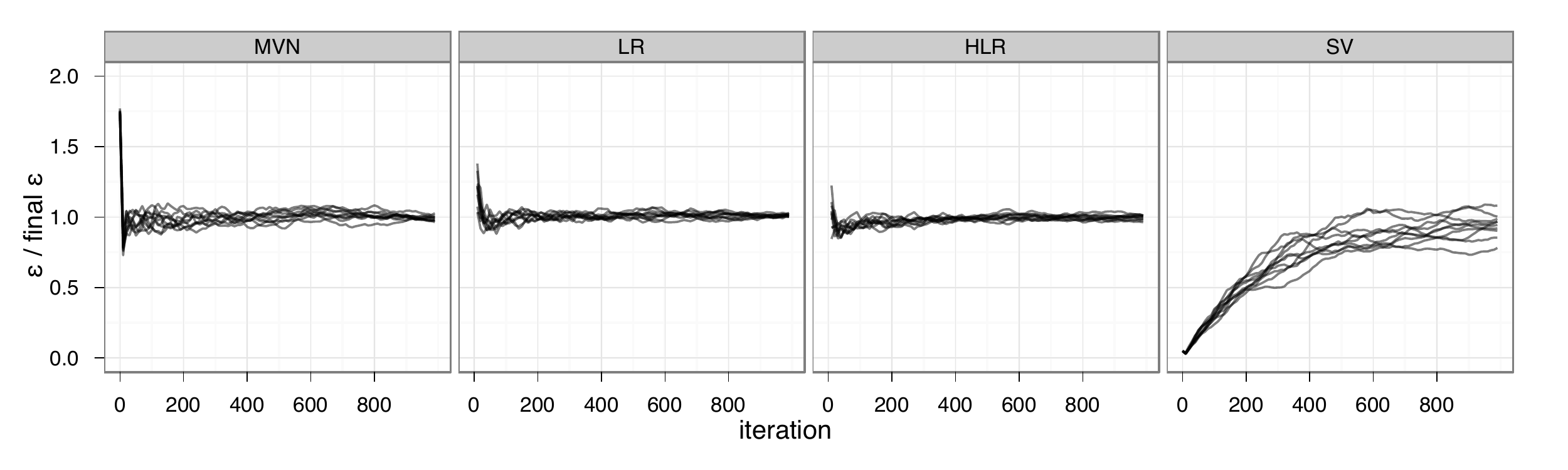}}
\end{center}
\vskip -0.3in
\caption{Plots of the convergence of $\bar\epsilon$ as a function of
  the number of iterations of NUTS with dual averaging with
  $\delta=0.65$ applied to the multivariate normal (MVN), logistic
  regression (LR), hierarchical logistic regression (HLR), and
  stochastic volatility (SV) models. Each trace is from an independent
  run. The y-axis shows the value of $\bar\epsilon$, divided by one of
  the final values of $\bar\epsilon$ so that the scale of the traces
  for each problem can be readily compared.}
\vskip -0.1in
\label{fig:epsilonconvergence}
\end{figure}

Figure \ref{fig:hconvergence} plots the realized versus target values
of the statistics $h^{\mathrm{HMC}}$ and $h^{\mathrm{NUTS}}$. The $h$
statistics were computed from the 1000 post-burn-in samples. The dual
averaging algorithm of section \ref{sec:sa} usually does a good job of
coercing the statistic $h$ to its desired value $\delta$.  It performs
somewhat worse for the stochastic volatility model, which we attribute
to the longer burn-in period needed for this model; since it takes
more samples to reach the stationary regime for the stochastic
volatility model, the adaptation algorithm has less time to tune
$\epsilon$ to be appropriate for the stationary distribution. This is
particularly true for HMC with small values of $\delta$, since the
overly high rejection rates caused by setting $\delta$ too small lead
to slower convergence.

\begin{figure}[t!]
\vskip-0.3in
\begin{center}
  \centerline{\includegraphics[width=1\columnwidth]{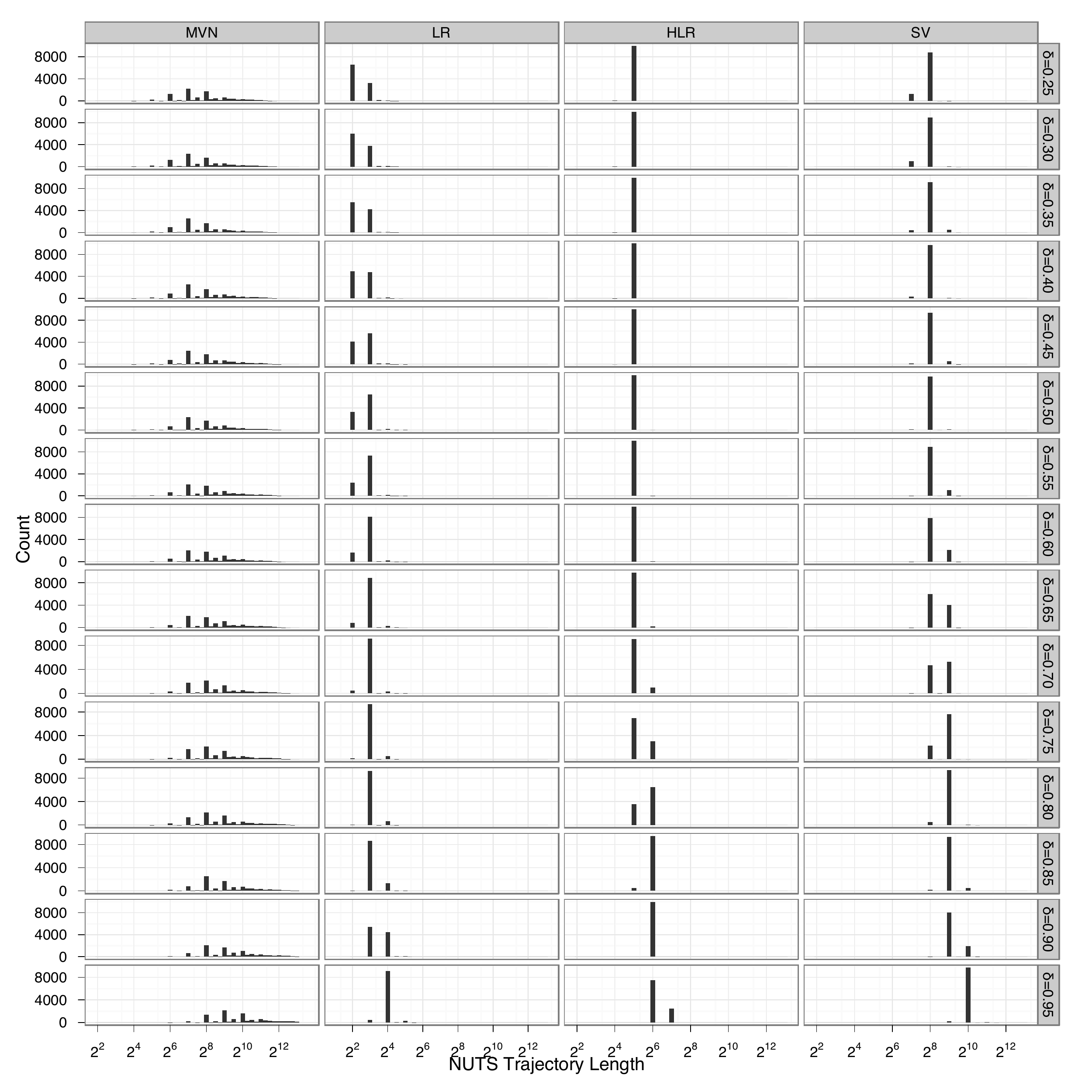}}
\end{center}
\vskip -0.3in
\caption{Histograms of the trajectory lengths generated by NUTS with
  various acceptance rate targets $\delta$ for the multivariate normal
  (MVN), logistic regression (LR), hierarchical logistic regression
  (HLR), and stochastic volatility (SV) models.}
\vskip -0.1in
\label{fig:nutstraj}
\end{figure}

Figure \ref{fig:epsilonconvergence} plots the convergence of the
averaged iterates $\bar\epsilon_m$ as a function of the number of dual
averaging updates for NUTS with $\delta=0.65$. Except for the
stochastic volatility model, which requires longer to burn in,
$\bar\epsilon$ roughly converges within a few hundred iterations.

\subsection{NUTS Trajectory Lengths}
\label{sec:nutstraj}
Figure \ref{fig:nutstraj} shows histograms of the trajectory lengths
generated by NUTS. Most of the trajectory lengths are integer powers
of two, indicating that the U-turn criterion in equation
\ref{eq:stopangle} is usually satisfied only after a doubling is
complete and not by one of the intermediate subtrees generated during
the doubling process. This behavior is desirable insofar as it means
that we only occasionally have to throw out entire half-trajectories
to satisfy detailed balance. 



\begin{figure}[t!]
\vskip-0.3in
\begin{center}
  \centerline{\includegraphics[width=1.0125\columnwidth]{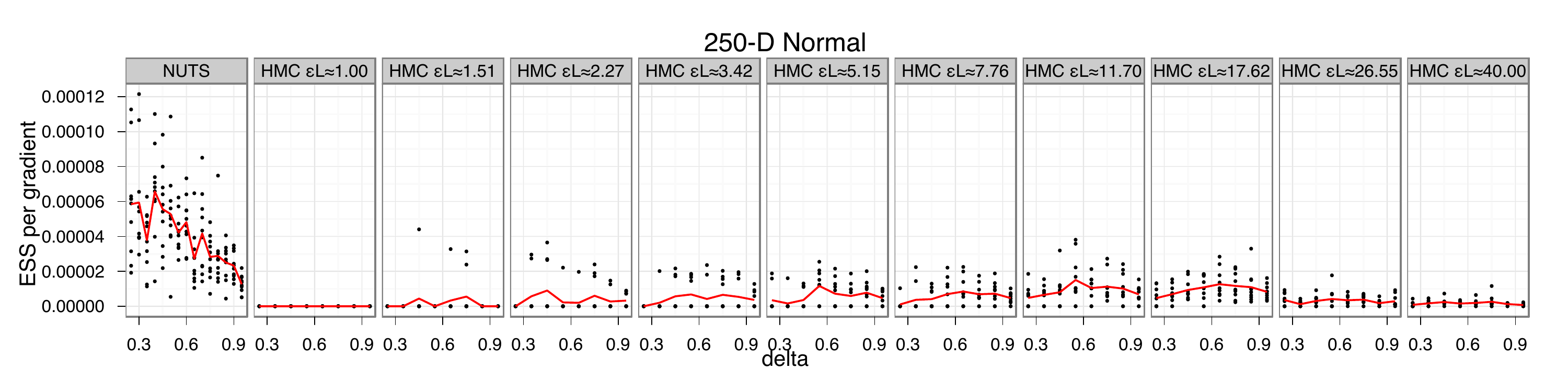}}
  \vskip-0.1in
  \centerline{\includegraphics[width=1.0125\columnwidth]{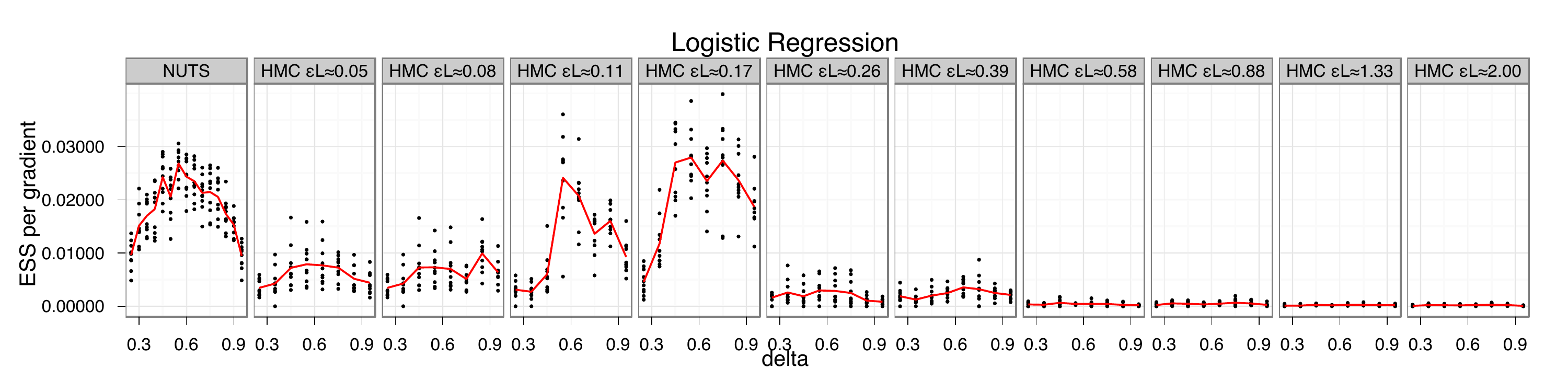}}
  \vskip-0.1in
  \centerline{\includegraphics[width=1.0125\columnwidth]{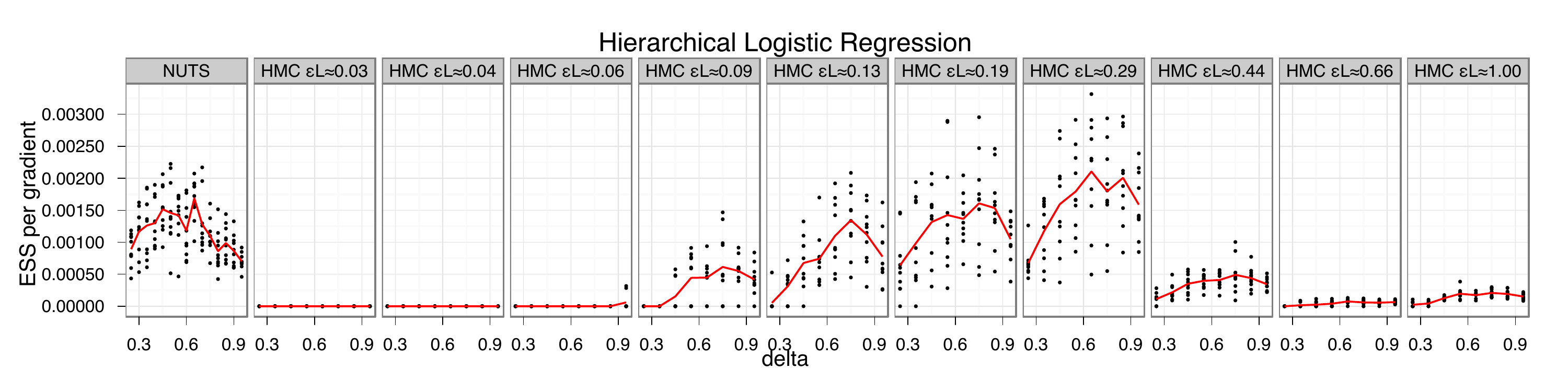}}
  \vskip-0.1in
  \centerline{\includegraphics[width=1.0125\columnwidth]{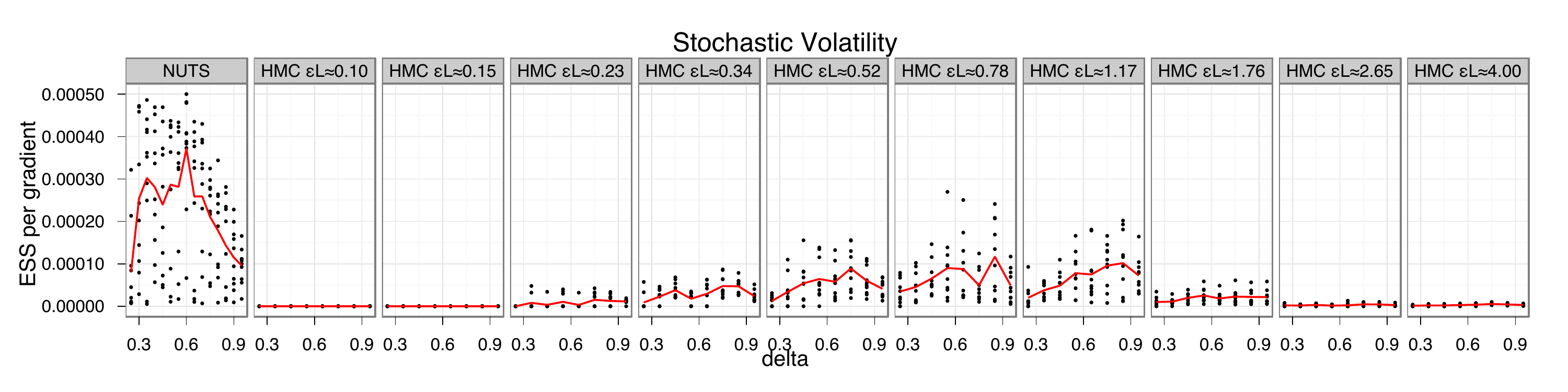}}
\end{center}
\vskip -0.3in
\caption{Effective sample size (ESS) as a function of $\delta$ and
  (for HMC) simulation length $\epsilon L$ for the multivariate
  normal, logistic regression, hierarchical logistic regression, and
  stochastic volatility models. Each point shows the ESS divided by
  the number of gradient evaluations for a separate experiment; lines
  denote the average of the points' y-values for a particular
  $\delta$. Leftmost plots are NUTS's performance, each other plot
  shows HMC's performance for a different setting of $\epsilon L$.}
\vskip -0.1in
\label{fig:ess}
\end{figure}

The trajectory length (measured in number of states visited) grows as
the acceptance rate target $\delta$ grows, which is to be expected
since a higher $\delta$ will lead to a smaller step size $\epsilon$,
which in turn will mean that more leapfrog steps are necessary before
the trajectory doubles back on itself and satisfies equation
\ref{eq:stopangle}.

\subsection{Comparing the Efficiency of HMC and NUTS}
\label{sec:hmcvnuts}


Figure \ref{fig:ess} compares the efficiency of HMC (with various
simulation lengths $\lambda\approx\epsilon L$) and NUTS (which chooses
simulation lengths automatically). The x-axis in each plot is the
target $\delta$ used by the dual averaging algorithm from section
\ref{sec:sa} to automatically tune the step size $\epsilon$.  The
y-axis is the effective sample size (ESS) generated by each sampler,
normalized by the number of gradient evaluations used in generating
the samples. HMC's best performance seems to occur around
$\delta=0.65$, suggesting that this is indeed a reasonable default
value for a variety of problems. NUTS's best performance seems to
occur around $\delta=0.6$, but does not seem to depend strongly on
$\delta$ within the range $\delta\in[0.45, 0.65]$. $\delta=0.6$
therefore seems like a reasonable default value for NUTS.

On the two logistic regression problems NUTS is able to produce
effectively independent samples about as efficiently as HMC can.  On
the multivariate normal and stochastic volatility problems, NUTS with
$\delta=0.6$ outperforms HMC's best ESS by about a factor of three.

As expected, HMC's performance degrades if an inappropriate simulation
length is chosen. Across the four target distributions we tested, the
best simulation lengths $\lambda$ for HMC varied by about a factor of
100, with the longest optimal $\lambda$ being 17.62 (for the
multivariate normal) and the shortest optimal $\lambda$ being 0.17
(for the simple logistic regression). In practice, finding a good
simulation length for HMC will usually require some number of
preliminary runs. The results in Figure \ref{fig:ess} suggest that
NUTS can generate samples at least as efficiently as HMC, even
discounting the cost of any preliminary runs needed to tune HMC's
simulation length.


\subsection{Qualitative Comparison of NUTS, Random-Walk Metropolis, and Gibbs}

\begin{figure}
\begin{center}
\vspace*{-8pt}
  \noindent\makebox[\textwidth]{
    \includegraphics[width=1.05\columnwidth]{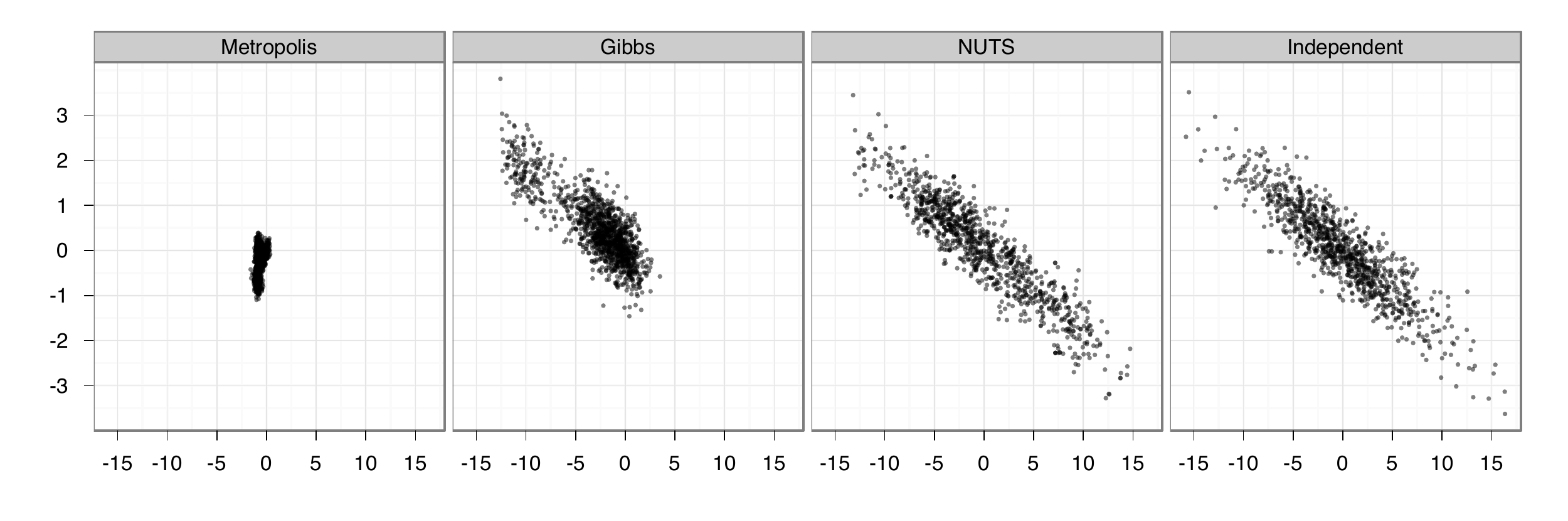}
  }
\end{center}
\vspace*{-12pt}
\caption{{\small\it Samples generated by random-walk Metropolis, Gibbs
    sampling, and NUTS.  The plots compare 1,000 independent draws
    from a highly correlated 250-dimensional distribution (right) with
    1,000,000 samples (thinned to 1,000 samples for display) generated
    by random-walk Metropolis (left), 1,000,000 samples (thinned to
    1,000 samples for display) generated by Gibbs sampling (second
    from left), and 1,000 samples generated by NUTS (second from
    right). Only the first two dimensions are shown here.
}}\label{fig:nuts-v-gibbs-rwm}
\end{figure}

In section \ref{sec:hmcvnuts}, we compared the efficiency of NUTS and
HMC. In this section, we informally demonstrate the advantages of NUTS
over the popular random-walk Metropolis (RWM) and Gibbs sampling
algorithms. We ran NUTS, RWM, and Gibbs sampling on the
250-dimensional multivariate normal distribution described in section
\ref{sec:models}.  NUTS was run with $\delta=0.5$ for 2,000
iterations, with the first 1,000 iterations being used as burn-in and
to adapt $\epsilon$.  This required about 1,000,000 gradient and
likelihood evaluations in total. We ran RWM for 1,000,000 iterations
with an isotropic normal proposal distribution whose variance was
selected beforehand to produce the theoretically optimal acceptance
rate of 0.234 \citep{Gelman:1996}. The cost per iteration of RWM is
effectively identical to the cost per gradient evaluation of NUTS, and
the two algorithms ran for about the same amount of time. We ran Gibbs
sampling for 1,000,000 sweeps over the 250 parameters. This took
longer to run than NUTS and RWM, since for the multivariate normal
each Gibbs sweep costs more than a single gradient evaluation; we
chose to nonetheless run the same number of Gibbs sweeps as RWM
iterations, since for some other models Gibbs sweeps can be done more
efficiently.

Figure \ref{fig:nuts-v-gibbs-rwm} visually compares independent
samples (projected onto the first two dimensions) from the target
distribution with samples generated by the three MCMC algorithms.  RWM
has barely begun to explore the space. Gibbs does better, but still
has left parts of the space unexplored.  NUTS, on the other hand, is
able to generate many effectively independent samples. 

We use this simple example to visualize the relative performance of
NUTS, Gibbs, and RWM on a moderately high-dimensional distribution
exhibiting strong correlations. For the multivariate normal, Gibbs or
RWM would of course work much better after an appropriate rotation of
the parameter space. But finding and applying an appropriate rotation
can be expensive when the number of parameters $D$ gets large, and RWM
and Gibbs both require $O(D^2)$ operations per effectively independent
sample even under the highly optimistic assumption that a
transformation can be found that renders all parameters i.i.d. and can
be applied cheaply (e.g. in $O(D)$ rather than the usual $O(D^2)$ cost
of matrix-vector multiplication and the $O(D^3)$ cost of matrix
inversion).  This is shown for RWM by \citet{Creutz:1988}, and for
Gibbs is the result of needing to apply a transformation requiring
$O(D)$ operations $D$ times per Gibbs sweep. For complicated models,
even more expensive transformations often cannot render the parameters
sufficiently independent to make RWM and Gibbs run efficiently. NUTS,
on the other hand, is able to efficiently sample from high-dimensional
target distributions without needing to be tuned to the shape of those
distributions.


\section{Discussion}
We have presented the No-U-Turn Sampler (NUTS), a variant of the
powerful Hamiltonian Monte Carlo (HMC) Markov chain Monte Carlo (MCMC)
algorithm that eliminates HMC's dependence on a number-of-steps
parameter $L$ but retains (and in some cases improves upon) HMC's
ability to generate effectively independent samples efficiently. We
also developed a method for automatically adapting the step size
parameter $\epsilon$ shared by NUTS and HMC via an adaptation of the
dual averaging algorithm of \citet{Nesterov:2009}, making it possible
to run NUTS with no hand tuning at all.  The dual averaging approach
we developed in this paper could also be applied to other MCMC
algorithms in place of more traditional adaptive MCMC approaches based
on the Robbins-Monro stochastic approximation algorithm
\citep{Andrieu:2008, Robbins:1951}.

In this paper we have only compared NUTS with the basic HMC algorithm,
and not its extensions, several of which are reviewed by
\citet{Neal:2011}. We only considered simple kinetic energy functions
of the form $\frac{1}{2}r\cdot r$, but both NUTS and HMC can benefit
from introducing a ``mass'' matrix $M$ and using the kinetic energy
function $\frac{1}{2}r^T M^{-1} r$. If $M^{-1}$ approximates the
covariance matrix of $p(\theta)$, then this kinetic energy function
will reduce the negative impacts strong correlations and bad scaling
have on the efficiency of both NUTS and HMC. Another extension of HMC
introduced by \citet{Neal:1994} considers windows of proposed states
rather than simply the state at the end of the trajectory to allow for
larger step sizes without sacrificing acceptance rates (at the expense
of introducing a window size parameter that must be tuned). The
effectiveness of the windowed HMC algorithm suggests that NUTS's lack
of a single accept/reject step may be responsible for some of its
performance gains over vanilla HMC.

\citet{Girolami:2011} recently introduced Riemannian Manifold
Hamiltonian Monte Carlo (RMHMC), a variant on HMC that simulates
Hamiltonian dynamics in Riemannian rather than Euclidean spaces,
effectively allowing for position-dependent mass matrices. Although
the worst-case $O(D^3)$ matrix inversion costs associated with this
algorithm often make it expensive to apply in high dimensions, when
these costs are not too onerous RMHMC's ability to adapt its kinetic
energy function makes it very efficient. There are no technical
obstacles that stand in the way of combining NUTS's ability to adapt
its trajectory lengths with RMHMC's ability to adapt its mass
matrices; exploring such a hybrid algorithm seems like a natural
direction for future research.

Like HMC, NUTS can only be used to resample unconstrained
continuous-valued variables with respect to which the target
distribution is differentiable almost everywhere. HMC and NUTS can
deal with simple constraints such as nonnegativity or restriction to
the simplex by an appropriate change of variable, but discrete
variables must either be summed out or handled by other algorithms
such as Gibbs sampling. In models with discrete variables, NUTS's
ability to automatically choose a trajectory length may make it more
effective than HMC when discrete variables are present, since it is
not tied to a single simulation length that may be appropriate for one
setting of the discrete variables but not for others.

Some models include hard constraints that are too complex to eliminate
by a simple change of variables. Such models will have regions of the
parameter space with 0 posterior probability. When HMC encounters such
a region, the best it can do is stop short and restart with a new
momentum vector, wasting any work done before violating the
constraints \citep{Neal:2011}. By contrast, when NUTS encounters a
0-probability region it stops short and samples from the set of points
visited up to that point, making at least some progress.

NUTS with dual averaging makes it possible for Bayesian data analysts
to obtain the efficiency of HMC without spending time and effort
hand-tuning HMC's parameters. This is desirable even for those
practitioners who have experience using and tuning HMC, but it is
especially valuable for those who lack this experience. In particular,
NUTS's ability to operate efficiently without user intervention makes
it well suited for use in generic inference engines in the mold of
BUGS \citep{Gilks:1992}, which until now have largely relied on much
less efficient algorithms such as Gibbs sampling. We are currently
developing an automatic Bayesian inference system called Stan, which
uses NUTS as its core inference algorithm for continuous-valued
parameters. Stan promises to be able to generate effectively
independent samples from complex models' posteriors orders of
magnitude faster than previous systems such as BUGS and JAGS.

In summary, NUTS makes it possible to efficiently perform Bayesian
posterior inference on a large class of complex, high-dimensional
models with minimal human intervention. It is our hope that NUTS will
allow researchers and data analysts to spend more time developing and
testing models and less time worrying about how to fit those models to
data.

\acks{This work was partially supported by Institute of Education
  Sciences grant ED-GRANTS-032309-005, Department of Energy grant
  DE-SC0002099, National Science Foundation grant ATM-0934516, and
  National Science Foundation grant SES-1023189.}

\appendix
\section{Estimating Effective Sample Size}
\label{app:ess}
For a function $f(\theta)$, a target distribution $p(\theta)$, and a
Markov chain Monte Carlo (MCMC) sampler that produces a set of $M$
correlated samples drawn from some distribution $q(\theta^{1:M})$ such
that $q(\theta^m)=p(\theta^m)$ for any $m\in\{1,\ldots,M\}$, the
effective sample size (ESS) of $\theta^{1:M}$ is the number of
independent samples that would be needed to obtain a Monte Carlo
estimate of the mean of $f$ with equal variance to the MCMC estimate
of the mean of $f$:
\begin{gather}
\label{eq:ess}
\nonumber \ESS_{q,f}(\theta^{1:M}) = M\frac{\V_q[\frac{1}{M}\sum_{s=1}^M f(\theta^s)]}
    {\frac{\V_p[f(\theta)]}{M}} = \frac{M}{1 + 2\sum_{s=1}^{M-1} (1-\frac{s}{M})\rho^f_s}; \\
\rho^f_s \equiv \frac{\Eq[(f(\theta^t)-\E_p[f(\theta)])(f(\theta^{t-s})-\E_p[f(\theta)])]}{\V_p[f(\theta)]},
\end{gather}
where $\rho^f_s$ denotes the autocorrelation under $q$ of $f$ at lag
$s$ and $\V_p[x]$ denotes the variance of a random variable $x$ under
the distribution $p(x)$.

To estimate ESS, we first compute the following estimate of the
autocorrelation spectrum for the function $f(\theta)$:
\begin{equation}
\hat\rho^f_s = \frac{1}{\hat\sigma^2_f(M-s)}
\sum_{m=s+1}^M (f(\theta^m)-\hat\mu_f)(f(\theta^{m-s})-\hat\mu_f),
\end{equation}
where the estimates $\hat\mu_f$ and $\hat\sigma^2_f$ of the mean and
variance of the function $f$ are computed with high precision from a
separated 50,000-sample run of NUTS with $\delta=0.5$. We do not take
these estimates from the chain whose autocorrelations we are trying to
estimate---doing so can lead to serious underestimates of the level of
autocorrelation (and thus a serious overestimate of the number of
effective samples) if the chain has not yet converged or has not yet
generated a fair number of effectively independent samples.

Any estimator of $\rho^f_s$ is necessarily noisy for large lags $s$,
so using the naive estimator $\hat\ESS_{q,f}(\theta^{1:M}) =
\frac{M}{1+2\sum_{s=1}^{M-1}(1-\frac{s}{M})\hat\rho^f_s}$ will yield
bad results. Instead, we truncate the sum over the autocorrelations
when the autocorrelations first dip below 0.05, yielding the estimator
\begin{equation}
\hat\ESS_{q,f}(\theta^{1:M}) = 
\frac{M}{1+2\sum_{s=1}^{M_f^{\mathrm{cutoff}}}(1-\frac{s}{M})\hat\rho^f_s}
;\quad M_f^\mathrm{cutoff} \equiv \min_s s \quad \textrm{s.t.}\ 
\hat\rho^f_s < 0.05.
\end{equation}

We found that this method for estimating ESS gave more reliable
confidence intervals for MCMC estimators than the autoregressive
approach used by CODA \citep{coda}. (The more accurate estimator comes
at the expense of needing to compute a costly high-quality estimate of
the true mean and variance of the target distribution.) The 0.05
cutoff is somewhat arbitrary; in our experiments we did not find the
results to be very sensitive to the precise value of this cutoff.

\bibliography{nutsarxiv.bbl}

\end{document}